\documentclass[%
reprint,
superscriptaddress,
amsmath,amssymb,
prb,
]{revtex4-1}
\usepackage{graphicx}
\usepackage{dcolumn}
\usepackage{bm}
\usepackage{hyperref}
\usepackage{xcolor}
\providecommand{\vect}[1]{{\bf {#1}}}
\providecommand{\uvect}[1]{{\bf \hat{#1}}}
\providecommand{\tensor}[1]{\bf \overset{\scriptsize$\leftrightarrow$}{#1}}
\providecommand{\myref}[1]{(\ref{#1})}
\allowdisplaybreaks
\begin{document}


\title{Probing and steering bulk and surface phonon polaritons in uniaxial materials using fast electrons: hexagonal boron nitride}

\author{C. Maciel-Escudero}
\affiliation{Materials Physics Center, CSIC-UPV/EHU, Donostia-San Sebasti\'{a}n, 20018, Spain}
\affiliation{CIC NanoGUNE BRTA and Department of Electricity and Electronics, EHU/UPV, Donostia-San Sebasti\'{a}n, 20018, Spain}
\author{Andrea Kone\v{c}n\'{a}}%
\affiliation{Materials Physics Center, CSIC-UPV/EHU, Donostia-San Sebasti\'{a}n, 20018, Spain}
\author{Rainer Hillenbrand}
\affiliation{CIC NanoGUNE BRTA and Department of Electricity and Electronics, EHU/UPV, Donostia-San Sebasti\'{a}n, 20018, Spain}
\affiliation{IKERBASQUE, Basque Foundation for Science, 48011 Bilbao, Spain
}
\author{Javier Aizpurua}
\affiliation{Materials Physics Center, CSIC-UPV/EHU, Donostia-San Sebasti\'{a}n, 20018, Spain}
\affiliation{Donostia International Physics Center DIPC, 20018 Donostia-San Sebasti\'{a}n, Spain}
\date{\today}
%
%
\begin{abstract}
\noindent We theoretically describe how fast electrons couple to polaritonic modes in uniaxial materials by analyzing the electron energy loss (EEL) spectra. We show that in the case of an uniaxial medium with hyperbolic dispersion, bulk and surface modes can be excited by a fast electron traveling through the volume or along an infinite interface between the material and vacuum. Interestingly, and in contrast to the excitations in isotropic materials, bulk modes can be excited by fast electrons traveling outside the uniaxial medium.  We demonstrate our findings with the representative uniaxial material hexagonal boron nitride. We show that the excitation of bulk and surface phonon polariton modes is strongly related to the electron velocity and highly dependent on the angle between the electron beam trajectory and the optical axis of the material. Our work provides a systematic study for understanding bulk and surface polaritons excited by a fast electron beam in hyperbolic materials and sets a way to steer and control the propagation of the polaritonic waves by changing the electron velocity and its direction.
\end{abstract}
%
%
\pacs{Valid PACS appear here}
\maketitle

%
%
\section{Introduction \label{sec1}}
Polar materials have become of high interest in the field of nanophotonics due to their ability to support phonon polaritons, quasi-particles which result from the coupling between electromagnetic waves and crystal lattice vibrations \cite{Zoo2016,Mills1974} with a characteristic wavelength lying in the mid-infrared region. These quasi-particles can enhance the electromagnetic field deep below the diffraction limit with large quality factors compared to infrared plasmons \cite{Hillenbrand2002,DeAbajo2016,Caldwell2019}, making them promising building blocks for infrared nanophotonics applications \cite{Zubin2014,Caldwell2015,Koppens2017,Caldwell2019-2}. 

One interesting 2D polar material is hexagonal boron nitride (h-BN) because of its high quality phonon polaritons and the easy preparation of the single atomic layers made by exfoliation \cite{Basov2014,Caldwell2014,Raschke2015,Hillenbrand2017,Walker2017,Ambrosio2018}. Besides being widely used in heterostructures \cite{Caldwell2015-2}, h-BN is emerging by itself as a versatile material offering novel optical and electro-optical functionalities. The crystal layer structure that constitues h-BN, mediated via van der Waals forces, produces an uniaxial optical response of the material. This implies that the dielectric function of h-BN needs to be described by a diagonal tensor $\tensor{\varepsilon}$ with two principal axes \cite{Caldwell2014,Raschke2015}: $\varepsilon_x=\varepsilon_y=\varepsilon_{\bot}$ and $\varepsilon_z=\varepsilon_{\parallel}$. When $\text{Re}(\varepsilon_{\parallel}) \cdot \text{Re}(\varepsilon_{\bot})<0$, phonon polaritons can propagate inside the material and exhibit a hyperbolic dispersion \cite{Kivshar2013,Zubin2014}, that is, the relationship between the different components of the polariton wavevector $\vect{k}(\omega)=(k_x,k_y,k_z)$ traces a surface in momentum space which corresponds to hyperboloids. For h-BN, one can find two energy bands (the Reststrahlen bands) where one of the principal components of the dielectric tensor is negative. Each Reststrahlen band is defined by the energy region between the transverse and longitudinal optical phonon energy, TO and LO, respectively ($\text{TO}_{\bot}$ and $\text{LO}_{\bot}$ for the upper Reststrahlen band and $\text{TO}_{\parallel}$ and $\text{LO}_{\parallel}$ for the lower Reststrahlen band, see Fig. 1). On the other hand, when $\text{Re}(\varepsilon_{\parallel}) \cdot \text{Re}(\varepsilon_{\bot})>0$, the isofrequency surfaces traced by the polariton wavevector in momentum space are ellipsoids. 
%
%
\begin{figure}[!ht]
\begin{center}
\includegraphics[scale=0.92]{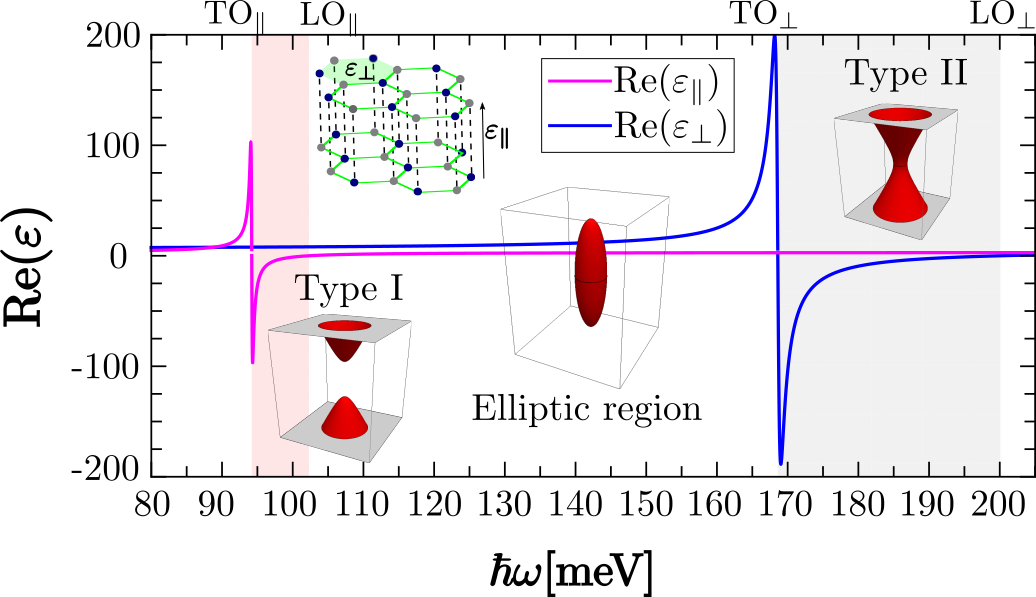}
\caption{Real parts of the components of the h-BN dielectric function. The shaded red area marks the lower Reststrahlen band and the grey area the upper Reststrahlen band. Insets illustrate the elliptic and the hyperbolic (Type I and Type II) isofrequency surfaces and the crystal lattice structure of h-BN.}
\label{fig1}
\end{center}
\end{figure}
%
\vspace{-5.2mm}

Figure \ref{fig1} depicts $\varepsilon_{\bot}$ and $\varepsilon_{\parallel}$ (see appendix \ref{appendA} for expressions and parameters of the dielectric tensor components), which represent the in-plane and out-of-plane dielectric components of h-BN, respectively. The energy range in Fig. \ref{fig1}, shaded in red, corresponds to the lower Reststrahlen band (94.2-102.3\,meV) where the real part of the out-of-plane permittivity is negative, leading to isofrequency surfaces in the form of two-sheet hyperboloids (inset Type I). The energy region shaded in grey corresponds to the upper Reststrahlen band (168.6-200.1\,meV), where the real part of the in-plane permittivity is negative and the isofrequency surfaces correspond to one-sheet hyperboloids (inset Type II).

Hyperbolic phonon polaritons excitable on h-BN within the range of 90 – 200 meV might be a key to many novel photonic technologies relying on the nanoscale confinement of light and its manipulation.  As a result, efficient design and utilization of h-BN structures require spectroscopic studies with adequate spatial resolution. This can be provided, for instance, by electron energy loss spectroscopy (EELS) using electrons as localized electromagnetic probes. Recently, instrumental improvements in EELS performed in a scanning transmission microscope (STEM-EELS) allowed to spatially map phonon polaritons \cite{,Crozier2014} and hyperbolic phonon polaritons in h-BN\cite{,Konecna2017,Ramasse2018}. The focused electron beam of an electron microscope has thus become a suitable probe to access the spectral information of low-energy excitations in technologically relevant materials, with nanoscale spatial resolution. Thus, EEL spectra in phononic materials can be of paramount importance to reveal the properties of phonon polariton excitations. 

In this work we first show that a fast electron traveling through bulk h-BN can excite volume phonon polaritons inside and outside the h-BN Reststrahlen bands. Our analysis reveals that the excitation of the volume polariton modes is strongly dependent on the electron velocity and also on the orientation between the electron beam trajectory and the h-BN optical axis. We then study the formation of wake patterns in the field distribution induced by the electron beam at h-BN. Our methodology allows us to connect the excitation of these wake fields with the different electron energy loss mechanisms experienced by the fast electron in the medium: (i) excitation of phonon polaritons or (ii) Cherenkov radiation. We also discuss the emergence of asymmetric wake patterns exhibited by the induced electromagnetic field when the electron beam trajectory sustains an angle relative to the h-BN optical axis. Finally, in the last two section of the paper we show that a fast electron beam interacting with a semi-infinite h-BN interface excite Dyakonov surface phonon polartions within the h-BN upper Reststrahlen band. We further demonstrate that the probing electron traveling above the h-BN in aloof trajectories excites volume phonon polaritons (remotely activation). All these findings offer a way to steer and control the propagation of the polaritonic waves and reveal the importance of the anisotropic optical response of the material in the EELS analysis.
%
%
\section{Excitation of infrared bulk modes in h-BN by a focused fast electron beam\label{sec2}}
%
%
%
\subsection{Bulk modes in h-BN}
According to Maxwell's equations in momentum-frequency ($k-\omega$) space, the dispersion relation for a wave propagating in the volume of an anisotropic material can be found from the following relationship: \cite{Eroglu2010}
\begin{equation}
\label{2.a}
\text{det}[\tensor{\bf G}^{-1}(\vect{k};\omega)]=\text{det} \left[\vect{k} \otimes \vect{k} - k^2 \tensor{\bf I} + k^2_0 \tensor{\varepsilon}\right]=0,
\end{equation}
where $\tensor{\bf G}^{-1}$ is the inverse of the Green's tensor, $\vect{k}(\omega)=(k_x,k_y,k_z)$ is the wavevector of the wave, $k_0=\omega/c$ is the magnitude of the wavevector in vacuum, $c$ is the speed of light, $\text{det}[\tensor{\bf x}]$ stands for the determinant of a matrix, $\otimes$ is the tensor product and $\tensor{\bf I}$ is the identity tensor. Particularly, for an uniaxial medium, the dielectric response can be described in tensor form as $\tensor{\varepsilon}(\omega)= \text{diag}[\varepsilon_{\bot},\varepsilon_{\bot},\varepsilon_{\parallel}]$. For this case, two solutions (modes) arise from Eq. \myref{2.a}, yielding the dispersion relation for ordinary waves
\begin{equation}
\label{2.b}
k^2= k^2_0 \varepsilon_{\bot},
\end{equation}
and the dispersion relation for extraordinary waves
\begin{equation}
\label{2.c}
\frac{k^2_x+k^2_y}{\varepsilon_{\parallel}}+\frac{k^2_z}{\varepsilon_{\bot}}=k^2_0.
\end{equation}
Equation \myref{2.b} represents concentric spheres in \textit{k}-space for a given energy $\hbar\omega$ (with $\varepsilon_{\bot}>0$), while Eq. \myref{2.c} represents hyperboloids or ellipsoids in the reciprocal space depending on the sign of the dielectric components $\varepsilon_{\parallel}$ and $\varepsilon_{\bot}$. Thus, we conclude that the isofrequency surfaces of the polariton wavevector $\vect{k}(\omega)$ in momentum space (for an uniaxial medium) represent geometrically spheres, ellipsoids or hyperboloids. For h-BN, the insets in Fig. \ref{fig1} depict the isofrequency surfaces for each energy region inside and outside the Reststrahlen bands.  
%
%
\subsection{Electron energy loss probability}
\begin{figure}[!h]
\begin{center}
\includegraphics[scale=0.70]{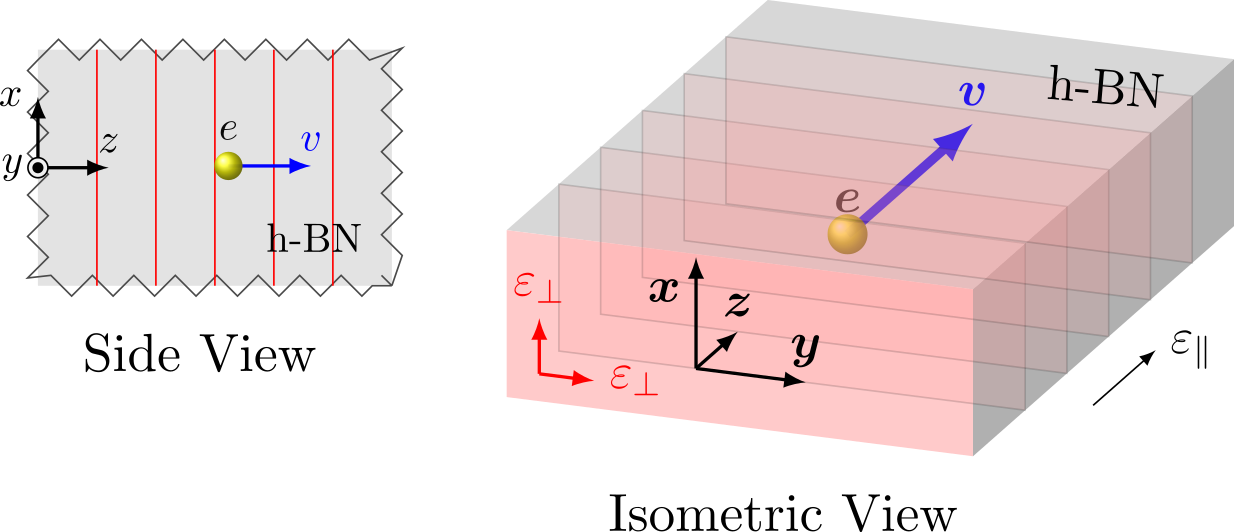}
\caption{Schematics of the electron traveling through the h-BN with velocity $\vect{v}=v \uvect{z}$ parallel to the h-BN optical axis (\textit{z}-direction).}
\label{fig2}
\end{center}
\end{figure}

Fast electron beams can couple to bulk polaritonic modes sustained in anisotropic media. We can observe this by analyzing the energy losses experienced by the electron when traveling in such media. Electron energy losses, $\Delta E_{\text{EELS}}$, can be calculated within classical electrodynamics as the work performed by the induced electromagnetic field, $\vect{E}^{\text{ind }}(\vect{r};t)$, on the probing electron\cite{Ritchie1957,Ritchie1985,DeAbajo2010-2,DeAbajo2019,Hohenester}
\begin{equation}
\label{2.d}
\Delta E_{\text{EELS}} = e \int  \text{d}\vect{r}_{e} \cdot \vect{E}^{\text{ind }}(\vect{r}_{e};t),
\end{equation}
where the integration is performed along the electron beam trajectory ${\vect{r}}_{e}(t)$, $e$ is the elementary charge, and $\vect{E}^{\text{ind}}(\vect{r};t)$ is evaluated along ${\vect{r}}_{e}(t)$. Notice that we approximate the electron beam as a classical point charge. The high-energy currents used in typical EELS experiments justify this approximation \cite{Richtie1981,Richtie1988,Rivacoba2014}. If we Fourier transform $\vect{E}^{\text{ind}}(\vect{r};t) \mapsto \vect{E}^{\text{ind}}(\vect{r};\omega)$ in Eq. \myref{2.d}, the electron energy losses can be written as
\begin{align}
\label{2.d.1}
\Delta E_{\text{EELS}} &= \frac{e}{2\pi} \int  \text{d}\vect{r}_{e} \cdot \int_{-\infty}^{\infty} \text{d}\omega\, \vect{E}^{\text{ind}}(\vect{r}_{e};\omega)\, e^{-i\omega t}\\
\nonumber
&=\int_0^{\infty} \text{d}\omega\, \int \text{d}L\, \hbar\omega \, \Gamma(\omega),
\end{align}
where one identifies the electron energy loss (EEL) probability per unit path, $\Gamma(\omega)$, as
\begin{equation}
\label{2.f}
\Gamma(\omega)= \frac{e}{\pi \hbar \omega} \text{Re} \left[ \vect{E}^{\text{ind}}({\vect{r}}_{e};\omega) \cdot \uvect{v} \, e^{-i\omega t_{e}} \right],
\end{equation}
with $\uvect{v}$ the unit vector in the same direction as the electron velocity $\vect{v}$ and $t_e$ is the time for the electron to travel a distance $\text{d}L$. Hence, to calculate $\Gamma(\omega)$ one needs to know the induced electric field $\vect{E}^{\text{ind}}(\vect{r};\omega)$. We derive below the expressions of the total electric field $\vect{E}^{\text{tot}}(\vect{r};\omega)$ and $\Gamma(\omega)$ for an electron beam trajectory parallel to the optical axis of h-BN, as depicted in Fig. \ref{fig2}.
%
%

It follows from Maxwell's equations that the field produced by the fast electron plus the induced electric field, namely the total electromagnetic field ($\vect{E}^{\text{tot}}(\vect{r};\omega)$) is given by
\begin{equation}
\label{2.g}
\vect{E}^{\text{tot}}(\vect{r};\omega)=-i \frac{\omega}{(2\pi)^3 c^2 \varepsilon_0} \int \text{d}^3\vect{k} \,\, \rho(\vect{k};\omega)\, \tensor{\bf G}(\vect{k};\omega) \cdot \vect{v}\, e^{i\vect{k} \cdot \vect{r}},
\end{equation} 
with $\varepsilon_{0}$ the vacuum permittivity and $\rho(\vect{k};\omega)= -2 e \pi \delta(\omega - \vect{k}\cdot\vect{v})$ the charge density of the probing electron. The integration in Eq. \myref{2.g} extends over the whole reciprocal space and the delta function introduced by the charge density assures conservation of energy and momentum. Indeed, one finds that in the non-relativistic limit the energy that the electron with initial velocity $\vect{v}$ transfers to the medium upon loosing momentum $\hbar\vect{k}$ is
\begin{equation}
\label{2.g.1}
\hbar \omega =\frac{(\vect{p}+\hbar \vect{k})^2}{2m}-\frac{p^2}{2m} = \hbar \vect{v}\cdot\vect{k}+\frac{\hbar^2}{2m}k^2,
\end{equation}
with $\vect{p}=m\vect{v}$ the initial momentum of the fast electron. By neglecting  recoil of the incident electron, from Eq. \myref{2.g.1} one arrives to the so-called non-recoil approximation where $\omega=\vect{k}\cdot\vect{v}$. Note that the \textit{z}-component of the wavevector is fixed by $k_z=\omega/v$ when the electron travels in the \textit{z}-direction.

To calculate the bulk loss probability $\Gamma^{\text{bulk}}(\omega)$  experienced by the  fast electron in the anisotropic medium we substitute Eq. \myref{2.g} into Eq. \myref{2.f}. Notice that a fast electron traveling in vacuum looses no energy, this allows to use $\vect{E}^{\text{tot}}(\vect{r};\omega)$ instead of $\vect{E}^{\text{ind}}(\vect{r};\omega)$ in Eq. \myref{2.f}. Using the cylindrical symmetry of the field produced by the fast electron one finds that 
\begin{equation}
\label{2.h}
\Gamma^{\text{bulk}}(\omega) = \int_{0}^{k^{\text{c}}_{\bot}} \text{d}k_{\bot} \,  P^{\text{bulk}}(k_{\bot};\omega),
\end{equation}
where
\begin{equation}
\label{2.i}
P^{\text{bulk}}(k_{\bot};\omega) = -\frac{2 e^2 k_{\bot} v}{(2\pi)^3 \hbar c^2 \varepsilon_0 v_z} \int_0^{2\pi} \text{d}\phi \, \text{Im} \left[\uvect{v} \cdot \tensor{\bf G}^* \cdot \uvect{v} \right],
\end{equation}
is the probability for the electron to transfer a transverse momentum $\hbar k_{\bot}$ (to the electron trajectory) upon lossing energy $\hbar\omega$. We will refer to this quantity as the momentum-resolved loss probability. In Eq. \myref{2.i} ${\tensor{\bf G}^*}={\tensor{\bf G}}(k_{\bot},\phi,k_z=\omega/v_z -  \vect{k}_{\bot}\cdot \vect{v}/v_z)$, and $\phi$ is the angle between $\vect{k}_{\bot}$ and the $k_x$-axis, with $\hbar k^{\text{c}}_{\bot}$ the maximum perpendicular momentum of the electrons selected by the collection aperture of the EELS spectrometer.

Particularly, when the electron beam trajectory points out in the same direction as the h-BN optical axis ($\vect{v}=v \uvect{z}$), expressions for $\vect{E}^{\text{tot}}(\vect{r};\omega)$ and thus for $\Gamma^{\text{bulk}}(\omega)$ can be found in a closed form (see appendix \ref{appendB} for the analytical formula of the Green's tensor in uniaxial anisotropic media):
\begin{equation}
\label{2.j}
\vect{E}^{\text{tot}}(\vect{r};\omega)= \frac{e}{2\pi \varepsilon_0} \frac{\omega}{v^2 \gamma_{\bot} \varepsilon_{\bot}} \vect{g}(\vect{r};\omega),
\end{equation}
where
\begin{align}
\label{2.k}
\nonumber
\vect{g}(\vect{r};\omega)&= e^{i\omega z/v} \left[  \frac{i}{\gamma_{\bot} }  K_0\left( \sqrt{\frac{\varepsilon_{\parallel}}{\varepsilon_{\bot}}} \frac{\lvert \omega \rvert}{\gamma_{\bot} v} R \right)  \uvect{z}  \right. \\
& \hspace{4mm} \left. -\, \text{sgn}(\omega) \sqrt{\varepsilon_{\parallel} \varepsilon_{\bot}}  K_1\left( \sqrt{\frac{\varepsilon_{\parallel}}{\varepsilon_{\bot}}} \frac{\lvert \omega \rvert}{\gamma_{\bot} v} R \right) \uvect{R} \right],
\end{align}
is written in cylindrical coordinates $\vect{r}=(\vect{R},z)=(x,y,z)$, $R=\sqrt{x^2+y^2}$, $K_0(x)$, $K_1(x)$ are the zero and first order modified Bessel functions of the second kind, sgn stands for the sign function and  $\gamma_{\bot}= 1/\sqrt{1-v^2 \varepsilon_{\bot}/c^2}$ is the Lorentz factor.
 
The momentum-resolved loss probability becomes
\begin{align}
\label{2.l}
\nonumber
P^{\text{bulk}}(k_{\bot};\omega) &= -\frac{2 e^2}{(2\pi)^2 \omega^2 \hbar \varepsilon_0} \text{Im} \left\{ \left[ k_0^2 \varepsilon_{\bot} - \frac{\omega^2}{v^2} \right]\right. \\
& \hspace{4mm} \left. \times \frac{k_{\bot}}{\varepsilon_{\parallel}[\varepsilon_{\bot} k_0^2-\omega^2/v^2] - \varepsilon_{\bot} k_{\bot}^2}  \right\},
\end{align}
and, substituting Eq. \myref{2.l} into Eq. \myref{2.h}, one finds that

\begin{align}
\label{2.m}
\nonumber
\Gamma^{\text{bulk}} (\omega) &=  \frac{e^2}{(2\pi)^2 \omega^2 \hbar \varepsilon_0} \text{Im} \left\{ \left[ k_0^2 - \frac{\omega^2}{\varepsilon_{\bot} v^2} \right] \right. \\
& \hspace{4mm} \left. \times\ln \left[\frac{\varepsilon_{\parallel} \varepsilon_{\bot} k_0^2- \varepsilon_{\parallel}\omega^2/v^2 - \varepsilon_{\bot} (k_{\bot}^c)^2}{\varepsilon_{\parallel} \varepsilon_{\bot} k_0^2 - \varepsilon_{\parallel} \omega^2/v^2} \right]  \right\}.
\end{align}
The non-retarded versions of $P^{\text{bulk}}(k_{\bot};\omega)$ and $\Gamma^{\text{bulk}}(\omega)$ can be obtained by setting $k_0$ equal to zero in Eqs. \myref{2.l} and \myref{2.m}. 

The spectrum of the momentum-resolved loss probability and the EEL probability provide valuable information which reveals the properties of the modes of the anisotropic material. We thus explore in the following the connection between the dispersion relation of the h-BN excitations in the upper Reststrahlen band with these two quantities.
%
%
%
\subsection{Upper Reststrahlen band}
%
%
%
\begin{figure*}[!ht]
\begin{center}
\includegraphics[scale=0.8]{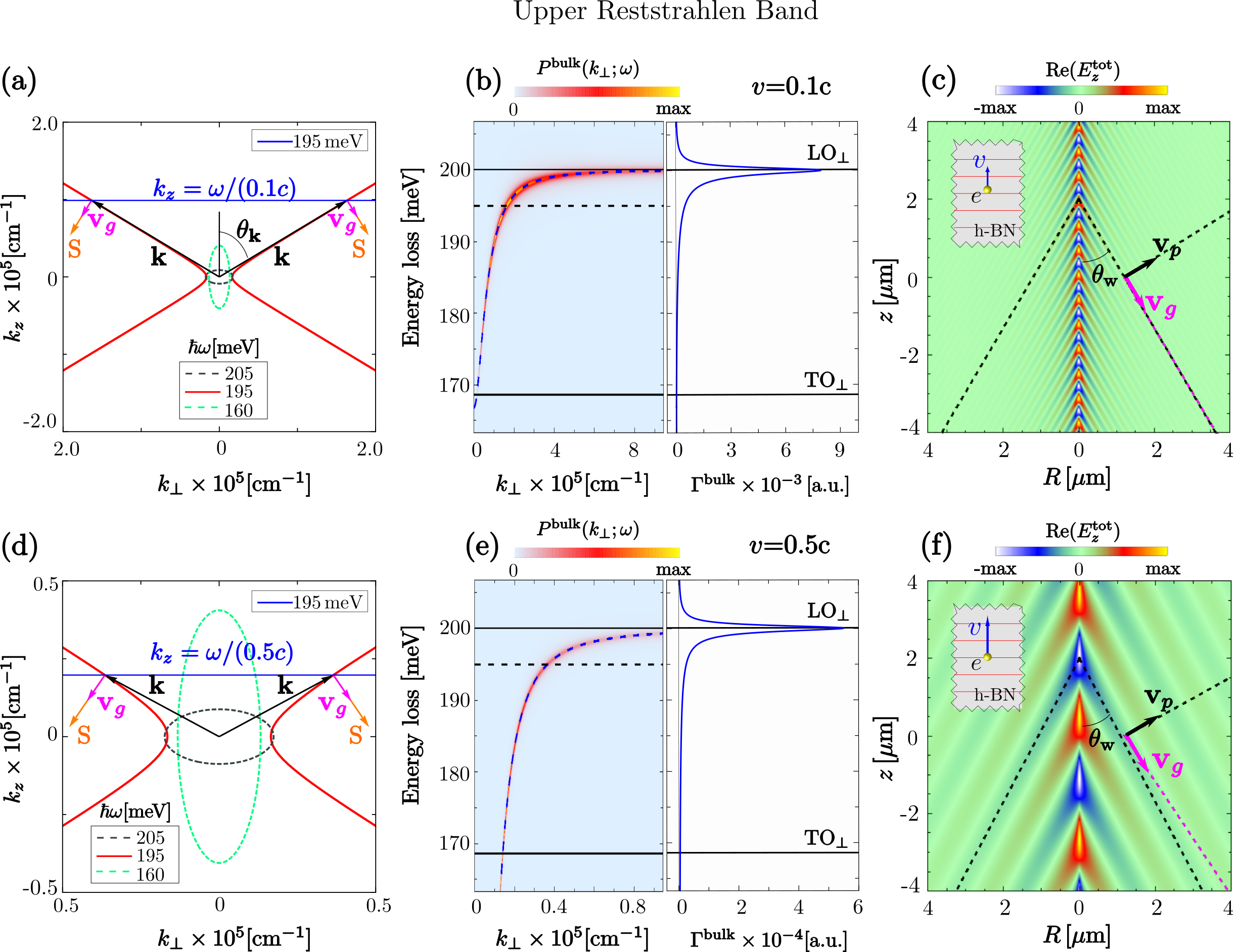}
\caption{(a) Isofrequency curves for energies inside (195\,meV, red solid line) and outside(160 and 205\,meV, green and black dashed lines) the upper Reststrahlen band plotted for the wavevector $k_z$ versus $k_{\bot}=\sqrt{k^2_x+k^2_y}$. The horizontal blue line represents the momentum $\hbar k_z=\hbar \omega/(0.1c)$ transferred by the fast electron to the polaritons when it travels along the \textit{z}-direction with $v=0.1c$. The blue line is evaluated at energy 195\,meV (caption at the top right of the figure). The black arrows represent the polariton wavevector $\vect{k}(\omega)$, $\theta_{\vect{k}}$ is the angle between $\vect{k}(\omega)$ and the $k_z$-axis, the magenta arrows represent the group velocity $\vect{v}_g$ and the orange arrows the Poynting vector $\vect{S}$. Panel (d) shows a zoom into (a). In (d) the horizontal blue line represents $k_z=\omega/(0.5c)$ . The contour plot (left panel) in (b) shows the momentum-resolved loss probability $P^{\text{bulk}}(k_{\bot};\omega)$ normalized to the maximum value (3\,a.u.) for $v=0.1c$. The right panel in (b) shows the energy loss probability $\Gamma^{\text{bulk}}(\omega)$ obtained by integrating $P^{\text{bulk}}(k_{\bot};\omega)$ over $k_{\bot}$ up to $k^c_{\bot}=0.05\,\mathring{\text{A}}^{-1}$. (e) same as in (b) but considering $v=0.5c$. (c) and (f) depict the real part of the \textit{z}-component of the total electric field induced by the fast electron along the cylindrical coordinates  ($R,z$) for the energy 195\,meV. The field plots are normalized to the maximum value in each case: (c) $1\times10^{-5}\,\text{a.u.}$ and (f) $7.5\times10^{-7}\,\text{a.u.}$ The insets  in (c) and (f) illustrate the electron beam trajectory and orientation of the h-BN crystal planes.} 
\label{fig3}
\end{center}
\end{figure*}
%

In the following we address the electron energy losses in h-BN  and the connection of these losses with the isofrequency surfaces of the material. We first show in Fig. \ref{fig3}a the isofrequency curve of a h-BN phonon polariton for an energy in the upper Reststrahlen band (red curve). We chose 195\,meV as a representative value of this band. When a fast electron beam is used to probe these excitations in the medium, the velocity of the electron determines the momentum transfer, as $\vect{k}\cdot\vect{v}=\omega$ (Eq. 8). If the electron is traveling along the \textit{z}-direction, then $k_z=\omega/v$ (blue horizontal line in Fig. \ref{fig3}a). Following Eq. \myref{2.c}, this also sets the value of the $\hbar k_{\bot}$ momentum component ($k^2_{\bot}=\varepsilon_{\parallel}k^2_0-\varepsilon_{\parallel} k^2_z/\varepsilon_{\bot}$) of the excited phonon polariton.

The intersections between $k_z=\omega/v$ and the isofrequency curves in the upper Reststrahlen band establish a relationship between the energy $\hbar\omega$ of the hyperbolic phonon polariton and its perpendicular momentum component $\hbar k_{\bot}$. In the left panel of Fig. \ref{fig3}b we plot this relationship (blue dashed line) and the momentum-resolved loss probability $P^{\text{bulk}}(k_{\bot};\omega)$ (light blue-yellow contour plot) for $v=0.1c$. We note that the highest values of $P^{\text{bulk}}(k_{\bot};\omega)$ coincide with the blue dashed line and its asymptotic behavior approaches $\text{LO}_{\bot}$ for large $k_{\bot}$. This demonstrates that electron energy losses in the upper band are due to phonon polariton excitations. We confirm this by integrating $P^{\text{bulk}}(k_{\bot};\omega)$ over $k_{\bot}$ up to a cutoff momentum $\hbar k^c_{\bot}$, which yields the EEL probability $\Gamma^{\text{bulk}}(\omega)$ (right panel of Fig. \ref{fig3}b). A clear peak can be observed at the longitudinal optical phonon. This energy loss peak is slightly asymmetric with a broader tail inside the Reststrahlen band compared to that outside the band.  Importantly, at energies above $\text{LO}_{\bot}$ no losses are found. This can be understood with the help of the isofrecuency curves in Fig. \ref{fig3}a. For instance, at energy 205\,meV (black dashed line, above the upper Reststrahlen band) the ellipse does not intersect the blue horizontal line and therefore there is no excitation above the upper band. For energies below $\text{TO}_{\bot}$, the ellipses may intersect or not the blue horizontal line of $k_z$ depending on the particular energy. For instance, at an energy of 160\,meV (green dashed line, below the upper Reststrahlen band in Fig. \ref{fig3}a) the ellipse does not cut $k_z=\omega/(0.1c)$ and therefore there is no excitation induced in that case. However, for lower energies the isofrequency surfaces can cut the $k_z$ line, and therefore an anisotropic dielectric mode can be excited (tail below 170\,meV in Fig. \ref{fig3}b). We learn from this analysis that the excitation of the phonon polariton modes close to the upper Reststrahlen band is highly dependent on the topology (hyperbolic or elliptic) of the isofrequency surfaces.

The dependency of phonon polaritons excitation on the isofrequency surface allows to control the polaritonic modes as we discuss now in Fig. \ref{fig3}c, where we show the real part of the \textit{z}-component of the total electric field at $\hbar\omega= 195\,\text{meV}$ (representing the energy within the hyperbolic dispersion regime in the upper Reststrahlen band), induced by a fast electron with velocity $v=0.1c$. A schematic representation of such electron beam trajectory is displayed in the inset of Fig. \ref{fig3}c.  We observe two important features: the formation of a wake pattern and an oscillatory behavior of the field in the \textit{z}-direction. This spatial periodicity is connected with the parallel momentum component ($\hbar k_z=\hbar \omega/v$) transferred by the electron since the observed wavelength along the \textit{z}-axis is $\lambda_z=2\pi/k_z$. This implies that the wavelength $\lambda_z$ decreases with increasing energy of the phonon polariton. Furthermore, the direction of the wake pattern is governed by the polariton phase velocity ($\vect{v}_p$ parallel to $\vect{k}(\omega)$, black arrow). The outward direction (relative to the electron beam trajectory) of the wavefronts is determined by the sign of the radial component of $\vect{v}_p$ relative to the radial component of the energy flow (given by the Pointing vector $\vect{S}=\vect{E}\times\vect{H}$ parallel to the group velocity $\vect{v}_g=\nabla_{\vect{k}} \omega$\cite{Felsen1994,Galyamin2011,Hillenbrand2015,Landau,Amnon}, magenta arrow). We recognize in Fig. \ref{fig3}a that the group and the phase velocities are nearly perpendicular, and their projection onto the radial axis are parallel, leading to a wave propagating away from the electron beam trajectory (positive phase and positive group velocity with respect to the energy propagation direction). It is worth noting that the projection of the group and the phase velocities onto the beam trajectory direction (\textit{z}-direction) leads to positive positive phase and negative group velocities relative to $S_z$ (Fig. \ref{fig3}a).

As pointed out, for each energy $\hbar\omega$, the velocity of the fast electron determines (primarily) the polariton wavevector parallel to the beam trajectory, $k_z$, and consequently the perpendicular wavevector $k_{\bot}$ (according to Eq. \myref{2.c}). To emphasize the velocity dependency, we perform the same analysis  (Fig. \ref{fig3}d-f) as in Fig. \ref{fig3}a-c but increasing the electron velocity to $v=0.5c$. In Fig. \ref{fig3}d a zoom into the isofrequency curve of Fig. \ref{fig3}a is presented, together with the value $k_z$ (horizontal blue line) determined by the electron velocity $v= 0.5c$. The increase of the electron velocity leads to the excitation of 195\,meV polaritons with reduced momentum (determined by the intersection of the blue horizontal line and the red isofrequency curve). By calculating the momentum-resolved loss probability $P^{\text{bulk}}(k_{\bot};\omega)$ (left panel of Fig. \ref{fig3}e) and the EEL probability $\Gamma^{\text{bulk}}(\omega)$ (right panel of Fig \ref{fig3}e) we find the same behavior as in Fig. \ref{fig3}b for $v=0.1c$, except for a one order of magnitude reduction in both $k_{\bot}$ and the value of the loss probability.

The differences in the properties of the phonon polaritons launched by the fast electron at both electron velocities are distinguishable in Fig. \ref{fig3}f, where we show the real part of  the \textit{z}-component of the total electric field induced by the fast electron with $v=0.5c$ at energy 195\,meV. The spatial period $\lambda_{z}$ of the polariton is longer compared to that in Fig. \ref{fig3}c as a result of the increase in the electron velocity (smaller $\hbar k_z$ transferred). Interestingly, the direction of the wake field is quite similar to that of panel \ref{fig3}c. This behavior is a specific feature of hyperbolic polaritons since the intersection of the blue line both for $v=0.1c$ and $v=0.5c$ occur at the asymptote of the hyperbola which results in polariton wavevectors that have very similar propagation direction but different absolute values.
 
%
%
\subsection{Lower Reststrahlen band}
\begin{figure*}[!ht]
\begin{center}
\includegraphics[scale=0.8]{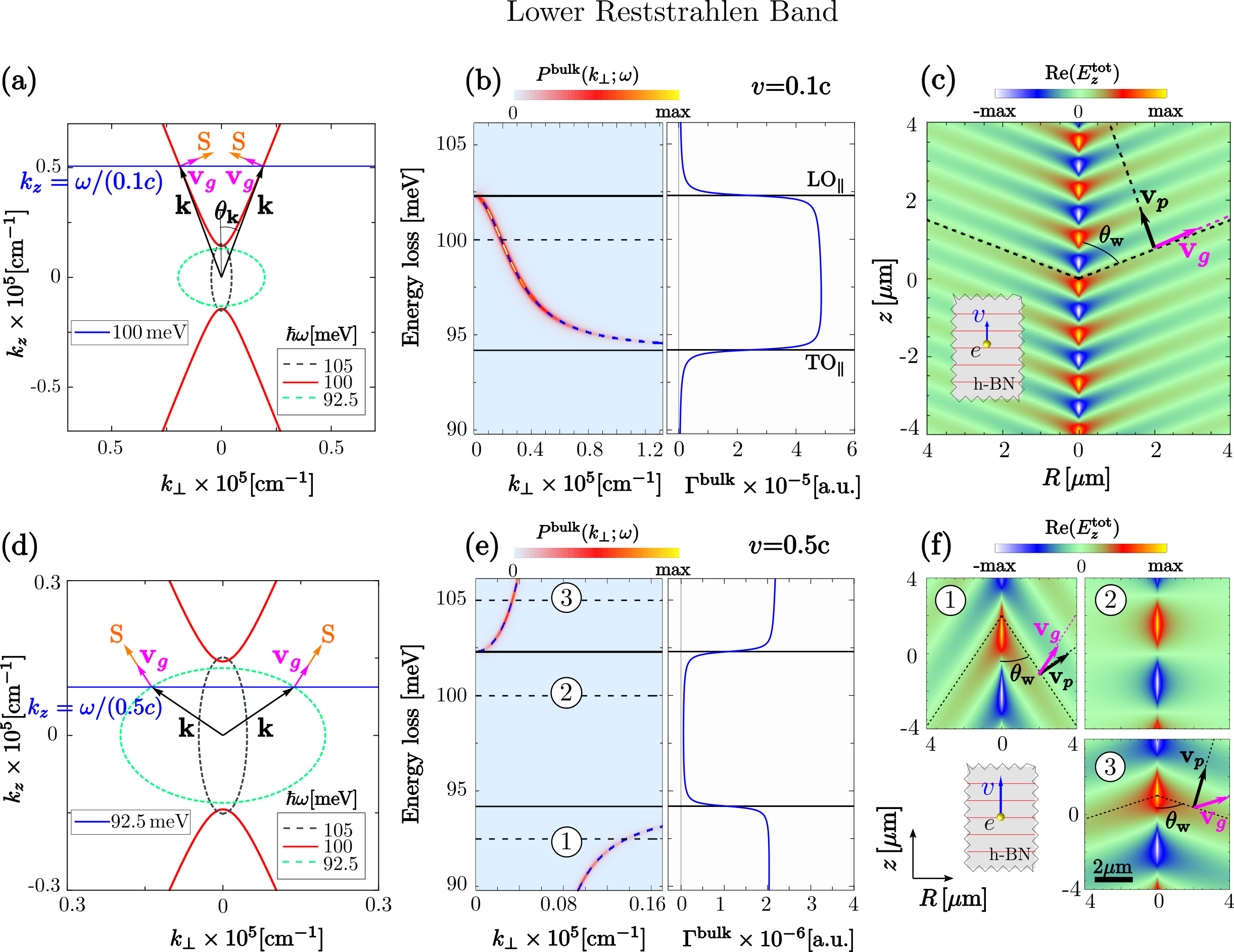}
\caption{(a) Isofrequency curves for energies inside (100\,meV, red solid line) and outside (92.5 and 105\,meV, green and black dashed lines) the lower Reststrahlen band plotted for the wavevector $k_z$ against $k_{\bot}=\sqrt{k^2_x+k^2_y}$. The horizontal blue line represents the momentum $\hbar k_z=\hbar \omega/(0.1c)$ transferred by the fast electron to the polaritons when it travels along the \textit{z}-direction with velocity $0.1c$. The blue line is evaluated at energy 100\,meV. The black arrows represent the polariton wavevector $\vect{k}(\omega)$, $\theta_{\vect{k}}$ is the angle between $\vect{k}(\omega)$ and the $k_z$-axis, the magenta arrows represent the group velocity $\vect{v}_g$ and the orange arrows the Poynting vector $\vect{S}$. Panel (d) shows a zoom into (a). In (d) the horizontal blue line represents $k_z=\omega/(0.5c)$ evaluated at energy 92.5\,meV. The contour plot (left panel) in (b) shows the momentum-resolved loss probability $P^{\text{bulk}}(k_{\bot};\omega)$ normalized to the maximum value (3\,a.u) for $v=0.1c$. The right panel in (b) shows the energy loss probability $\Gamma^{\text{bulk}}(\omega)$ obtained by integrating $P^{\text{bulk}}(k_{\bot};\omega)$ over $k_{\bot}$ up to $k^c_{\bot}=0.05\,\mathring{\text{A}}^{-1}$. (e) same as in (b) but considering $v=0.5c$. For this case the maximum of $P^{\text{bulk}}(k_{\bot};\omega)$ is 2.5\,a.u. Panels (c) and (f) depict the real part of the \textit{z}-component of the total electric field induced by the fast electron along the cylindrical coordinates $(R,z)$ for the energies: (c) 100\,meV (for $v=0.1c$)  and (f) 92.5, 100 and 105\,meV (for $v=0.5 c$). The field plots are normalized to the maximum value in each case: (c) $1\times10^{-6}\,\text{a.u.}$ and  (f) $7.5\times10^{-8}\,\text{a.u.}$. The insets in (c) and (f) illustrate the electron beam trajectory.}
\label{fig4}
\end{center}
\end{figure*}

In Fig. \ref{fig4}a we show the isofrequency curve of h-BN phonon polaritons for an energy in the lower Reststrahlen band (red line). Note that the hyperbolas are rotated by $90^{\circ}$ as compared to the upper Reststrahlen band (see Fig. \ref{fig3}a and \ref{fig3}d). However, the momentum $\hbar k_z$ transferred by the fast electron to the phonon polaritons is still given by the crossing of the hyperbolas with the horizontal blue line (representing $k_z =\omega/v$ for $v=0.1c$ in Fig. \ref{fig4}a). From Eq. \myref{2.c} we obtain the polariton perpendicular momentum $\hbar k_{\bot}$, which is shown in Fig. \ref{fig4}b as a function of energy $\hbar\omega$ (dashed blue curve). We also plot the momentum-resolved loss probability $P^{\text{bulk}}(k_{\bot};\omega)$ for energies within the lower Reststrahlen band. Notice that the highest values of $P^{\text{bulk}}(k_{\bot};\omega)$ (red and yellow colors in the contour plot) coincide perfectly with the blue dashed curve, demonstrating that the electron energy losses in the lower band are also governed by polariton excitations. However, in contrast to the upper band, we find that the dashed blue curve has a negative slope, $\text{d}\omega/\text{d}k_{\bot} < 0$, indicating that the group and the phase velocities are antiparallel (have opposite sign) along the radial direction. We will show below with Fig. \ref{fig4}c that the phase velocity in the radial direction is indeed antiparallel (negative) relative to the Poynting vector (energy flow) while the group velocity in the radial direction is parallel (positive), which is a consequence of the phase and group velocity vectors being perpendicular to each other and rotated by $90^{\circ}$ degrees compared to the upper Reststrahlen band.
%
%
%
%

To obtain spectroscopic information on the excitations in the lower Reststrahlen band, we calculate the EEL probability $\Gamma^{\text{bulk}}(\omega)$ by integration of $P^{\text{bulk}}(k_{\bot};\omega)$ in momentum space (right panel in Fig. \ref{fig4}(b)). Contrary to the upper Reststrahlen band, we observe a uniform and relatively small loss probability between $\text{TO}_{\parallel}$ and $\text{LO}_{\parallel}$ without the appearance of a sharp peak around $\text{LO}_{\parallel}$. We explain this finding by: 1) the large cutoff momenta ($\hbar k^c_{\bot}$) imposed by the aperture of the microscope detector and 2) the relationship between the energy and the transverse momentum of the polaritons in the lower band (see Fig. \ref{fig4}b left panel). Indeed, we observe in Fig. \ref{fig4}b that the asymptotic behavior of the blue dashed line tends to $\text{TO}_{\parallel}$ for large $k_{\bot}$. This shows that low energy hyperbolic phonon polaritons, close to $\text{TO}_{\parallel}$, largely contribute to the energy losses for large $k^c_{\bot}$ values. Contrary to the upper band, where the high momenta contribution to the electron energy losses comes from polaritons with high energy, close to $\text{LO}_{\bot}$ (Fig. \ref{fig3}b, left panel).  We address the reader to appendix \ref{appendC} where we show $\Gamma^{\text{bulk}}(\omega)$ in the lower Reststrahlen band for different cutoff values.

The excitation of phonon polaritons (within the lower Reststrahlen band) by the probing electron can be observed in Fig. \ref{fig4}c, where we show the real part of the \textit{z}-component of the total electric field induced at energy $\hbar\omega=100\,\text{meV}$. Analogously to the upper band, the oscillatory behavior of the field distribution along the \textit{z}-direction is governed by the transferred momentum $\hbar k_z$. Interestingly, the wake pattern is reversed compared to that for the upper Reststrahlen band (Figs. \ref{fig3}c and \ref{fig3}f), i.e., the wavefronts are propagating toward the electron beam \cite{Galyamin2011,Tao2016,Tao2019}. By plotting the group and phase velocity vectors onto the field plots (purple and black arrows, respectively; also plotted in Fig. \ref{fig4}a), we can clearly recognize that the projections of both vectors onto the radial axis (perpendicular to the electron beam trajectory) are antiparallel. This leads to a negative phase and positive group velocity relative to the Poynting vector direction (which points always away from the electron beam to preserve causality) along the radial axis. The negative phase velocity in the radial direction is a direct result of the phase velocity vector being nearly pependicular to the Poynting vector, both being rotated by $90^{\circ}$ as compared to the upper Reststrahlen band (where both phase and group velocities are positive relative to energy propagation in the radial direction, see Figs. \ref{fig3}c and \ref{fig3}f).

When the velocity of the electron is increased up to 50\,\% the speed of light, the $k_z$ component of the wavevector parallel to the beam trajectory is reduced. In this case, the matching between the red hyperbola and the horizontal blue line is prevented as observed in Fig. \ref{fig4}d. This mismatch of energy and momentum forbids the excitation of hyperbolic phonon polaritons. However, the blue line intersects the elliptical isofrequency surface of anisotropic bulk phonon polaritons (dielectric) above and below the lower Reststrahlen band (black and green dashed curves calculated for 105 and 92.5\,meV, respectively). The matching of energy and momentum at the intersections of the elliptical isofrequency surfaces leads to the excitation of the dielectric modes, as demonstrated by calculating the momentum-resolved loss probability $P^{\text{bulk}}(k_{\bot};\omega)$ (left panel of Fig. 4e). This loss probability is determined by the relationship between the energy of the elliptical polartions and the perpendicular momentum component (dashed blue lines, showing $\omega(k_{\bot})$ of the elliptical polaritons). The integration of $P^{\text{bulk}}(k_{\bot};\omega)$ in the reciprocal space subsequently yields small energy loss probabilities outside the Reststrahlen band, whereas inside the Reststrahlen band the loss probability is negligible due to absence of polariton excitations.

In Fig. \ref{fig4}f we show the total electric field induced by the electron beam for energies inside (marked 2) and outside (marked 1 and 3) the lower Reststrahlen band. We can observe the formation of wake patterns only for those energies where the dielectric modes are excited (marked as 1 and 3). Importantly, the wake wavefronts propagate outward the beam trajectory as a consequence of the group ($\vect{v}_g$) and phase velocities ($\vect{v}_p$) being parallel (positive) relative to the Poyting vector in the radial direction (Fig. \ref{fig4}d). We can also notice that the projection of these velocity vectors onto the \textit{z}-direction is positive. This demonstrates that the radial and \textit{z} projections of $\vect{v}_p$ and $\vect{v}_g$ for elliptical polaritons are positive, contrary to the hyperbolic regime (Reststrahlen bands) where one of the components is negative (Figs. \ref{fig3}c,f and \ref{fig4}c).
%
%
%
\subsection{Induced wake patterns and Cherenkov radiation}
We have shown in sections \ref{sec2}C (Figs. \ref{fig3}c,f) and  \ref{sec2}D (Figs. \ref{fig4}c,f,) that the field distributions produced by a fast electron traveling through h-BN can exhibit wake patterns. The excitation of these patterns (for energies inside and outside the Reststrahlen bands) is connected to the different mechanisms of energy losses experienced by the fast electron in the h-BN. In the following we discuss this connection.

First, it is worth noting that the excitation of the wake fields inside the Reststrahlen bands occurs for energies where electron losses appear (compare Fig. \ref{fig4}c with the image in Fig. \ref{fig4}f labeled as 2). As we pointed out, the electron energy losses within the Reststrahlen bands correspond to the excitation of hyperbolic phonon polaritons. This implies that the wake fields are associated to the excitation of coherent-charge density fluctuations\cite{Ritchie1974,Echenique1979,Echenique1979-2,DeAbajo1992,Liu2017,Tao2019} in the h-BN, namely, the phonon polaritons. 

In contrast to the wake fields inside the Reststrahlen bands, the emergence of the wake patterns outside the bands (see Fig. \ref{fig4}, images labeled as 1 and 3) occurs due to a different physical process to that of the excitation of hyperbolic phonon polaritons. Outside the Reststrahlen bands the h-BN dielectric function is purely dielectric and thus the electron energy losses correspond to the radiation emitted by the electron when it passes through the medium with velocity larger than the speed of light (in the h-BN). This mechanism is known as Vavilov-Cherenkov radiation\cite{Cherenkov1934,Cherenkov1937,Tamm1937,Tamm1939,Lucas1970,Silcox1979,Ginzburg1996}. We have confirmed that the losses in this energy range are present even in the absence of damping in the material (not shown), confirming that the losses are due to Cherenkov radiation in this case. This only happens for velocities which fulfill
\begin{equation}
\label{2.o}
v>\frac{c}{\sqrt{\varepsilon_{\bot}}}. 
\end{equation}
being consistent with the condition for excitation of Cherenkov radiation \cite{Rabernik2013}. 

Finally, one can also note that the excitation of the wake fields in the lower Reststrahlen band depends on the electron velocity (compare Figs. \ref{fig4}c and \ref{fig4}f label 2). Indeed, for energies in the lower band one can deduce from Eq. \myref{2.j} that the wake patterns appear under the following condition:
\begin{equation}
\label{2.n}
\frac{\varepsilon_{\parallel}}{\varepsilon_{\bot}}<\frac{v^2}{c^2}\varepsilon_{\parallel} \hspace{2mm} \text{or} \hspace{2mm} v<\frac{c}{\sqrt{\varepsilon_{\bot}}},
\end{equation}
where only the real part of the dielectric function is considered. Interestingly, one can observe that the velocity of the fast electron fulfills different conditions for different energy ranges (compare Eqs. \myref{2.o} and \myref{2.n}). This difference is a direct consequence of the distinct physical processes in the excitation of the wake fields. 

The different nature of the excitation of the wake fields outside and inside the Reststrahlen bands is also reflected in the angle $\theta_{\text{w}}=90^{\circ}-\theta_{\vect{k}}$ that the wake patterns sustain with respect to the electron beam trajectory. An analysis of this angle and its relationship with Eqs. \myref{2.o} and \myref{2.n} is developed at the end of appendix \ref{appendD}.
%

%
%
\subsection{Asymmetric wake patterns induced by tilting the electron beam trajectory}
As we pointed out in the last sections, the excitation of hyperbolic phonon polaritons can be controlled by the velocity of the fast electrons. In the following we study how steering of phonon polaritons can be controlled via the angle $\alpha$ between the electron beam trajectory and the h-BN optical axis.
%
%
\begin{figure}[!ht]
\begin{center}
\includegraphics[scale=0.70]{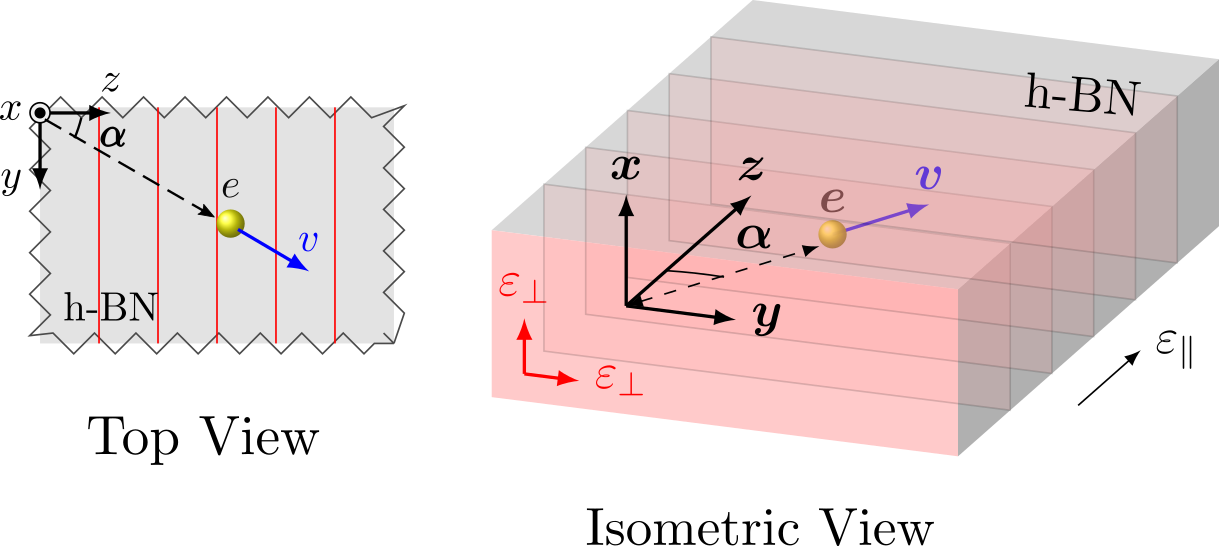}
\caption{Schematics of the electron traveling through the h-BN with velocity $\vect{v}=v(0,\sin\alpha,\cos\alpha)$ at an angle $\alpha$ with respect to the optical (\textit{z}-) axis of h-BN.}
\label{fig5}
\end{center}
\end{figure}

\begin{figure*}[!ht]
\begin{center}
\includegraphics[scale=0.77]{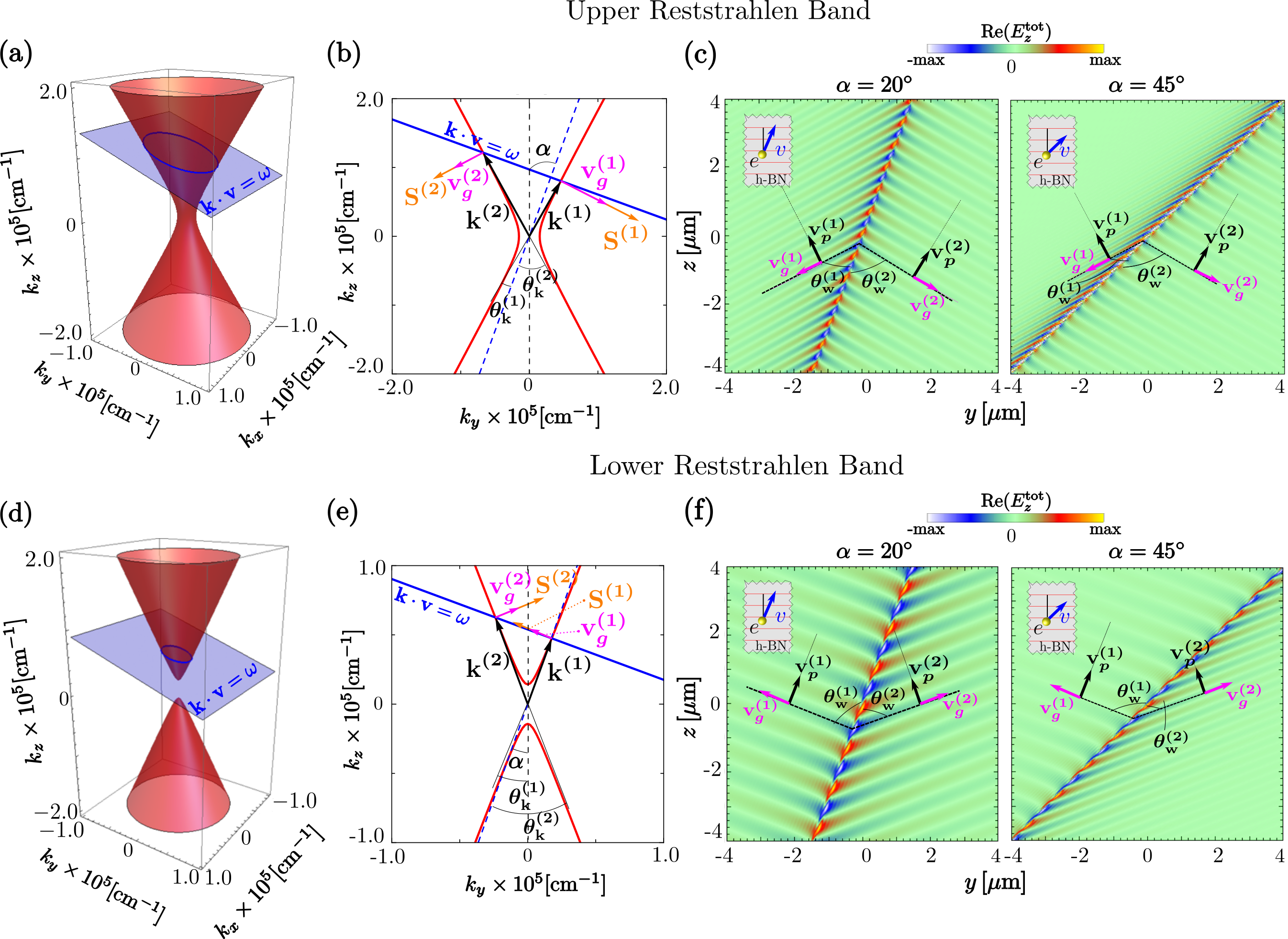}
\caption{Isofrequency surfaces for (a) the upper and (d) the lower Reststrahlen bands for representative energies in each band: (a) 180\,meV and (d) 100\,meV. The blue inclined plane represents the condition for the conservation of energy and momentum in the non-recoil approximation: $\vect{k}\cdot\vect{v}=\omega$ for an electron with $v=0.1c$ and $\alpha=20^{\circ}$. Panels (b) and (e) next to each hyperboloid depict the intersection between the blue plane and the hyperboloids in the $k_yk_z$-plane. In these projections, the black arrows represent the two wavevector solutions $\vect{k}^{(1)}$, $\vect{k}^{(2)}$ with angles $\theta^{(1)}_{\vect{k}}$,  $\theta^{(2)}_{\vect{k}}$ with respect to the beam trajectory (blue dashed line), the magenta arrows represent the group velocities $\vect{v}^{(1)}_{g}$, $\vect{v}^{(2)}_{g}$ and the orange arrows the Poynting vectors $\vect{S}^{(1)}$, $\vect{S}^{(2)}$. The contour plots in (c) and (f) show the normalized real part of the \textit{z}-component of the total electric field in the $yz$-plane for the energies: (c) 180\,meV and (f) 100\,meV. We plot the field distributions for two different angles of the electron beam trajectory: $20^{\circ}$ (left panels) and $45^{\circ}$ (right panels). The maximum values of the field plots are: (c) $4\times 10^{-6}\,\text{a.u.}$ and (f) $>1.5\times 10^{-6}\,\text{a.u.}$}
\label{fig6}
\end{center}
\end{figure*}

When the electron travels at an angle $\alpha$ relative to the h-BN optical axis (illustrated in Fig. \ref{fig5}), the condition for the conservation of energy and momentum given by Eq. \myref{2.g.1} ($\vect{k}\cdot\vect{v}=k_y v_y+ k_z v_z =\omega$) is represented by an inclined plane in momentum space (blue planes in Figs. \ref{fig6}a and \ref{fig6}d). The magnitude of the momentum transferred by the electron to the phonon polaritons (along the beam trajectory given by $\uvect{v}$) is still given by $\hbar k_{\uvect{v}}=\hbar \omega/v$. The polariton wavevector can be obtained from the intersection between the blue plane $\vect{k}\cdot\vect{v}=\omega$ and the isofrequency surfaces (red hyperboloids in Figs. \ref{fig6}a and \ref{fig6}d). Interestingly, we observe that the intersections are not cylindrically symmetric with respect to the \textit{z}-axis (Figs. \ref{fig6}a,c). This implies that the polaritonic wave will propagate asymmetrically with respect to the electron beam trajectory. Indeed, depending on the direction of propagation, the intersection between the blue planes and the red hyperboloids in Figs. \ref{fig6}a,d will occur at wavevectors $\vect{k}^{(1)}$ and $\vect{k}^{(2)}$ whose \textit{z}-component can be the same (symmetrical case) or different (asymmetrical case). To better understand the different  asymmetries in the propagation of the polaritonic wave we refer the reader to appendix \ref{appendD}, where we show the intersection of the blue plane and the red hyperboloids (Figs. \ref{fig6}a,d) for selected directions of the wavevector. Notice that the symmetric case is similar to the one we discussed in sections \ref{sec2}C and \ref{sec2}D. Therefore we will focus here on the analysis of the polariton propagation direction which shows the largest asymmetry, that is, the $k_yk_z$-plane.
%
%
%

We show in Fig. \ref{fig6}a the plane $\vect{k}\cdot\vect{v}=\omega$ for $v=0.1c$ (blue surface) and the isofrequency hyperboloid (red surface) for a representative energy in the upper Reststrahlen band ($\hbar\omega=180\,\text{meV}$). In Fig. \ref{fig6}b, we plot the projection of the intersection between the blue plane and the red hyperboloid in the $k_yk_z$-plane. The blue dashed line represents the electron beam trajectory and the black dashed line the $k_{z}$-axis. One can notice that the matching between the blue solid line and the red hyperbola (Fig. \ref{fig6}b) occurs at wavevectors $\vect{k}^{(1)}$ and $\vect{k}^{(2)}$ whose \textit{z}-component is different. Thus, the projections onto the \textit{z}-axis of the phase velocities $\vect{v}^{(1)}_{p}$ and $\vect{v}^{(2)}_{p}$ (parallel to $\vect{k}^{(1)}$ and $\vect{k}^{(2)}$, black arrows) are also different. Due to the hyperbolic shape of the isofrequency curve the \textit{z}-component of the group velocities $\vect{v}^{(1)}_{g}$ (parallel to the Poynting vector $\vect{S}^{(1)}$, right orange arrow in Fig. \ref{fig6}b) and $\vect{v}^{(2)}_{g}$ (parallel to the Poynting vector $\vect{S}^{(2)}$, left orange arrow in Fig. \ref{fig6}b) are also asymmetric. This difference (asymmetry) in the components of the two phase and group velocities leads to a highly asymmetric propagation of the polaritonic wave with respect to the electron beam trajectory.

The dependency of the polaritonic waves on the angle $\alpha$ can be observed in Fig. \ref{fig6}c, where we plot the real part of the \textit{z}-component of the total electric field produced by the fast electron at energy $\hbar\omega=180\,\text{meV}$ and $v=0.1c$ when $\alpha=20^{\circ}$ and $\alpha=45^{\circ}$. Similar to the parallel trajectory (sections \ref{sec2}C and \ref{sec2}D), one can notice the formation of wake patterns and the spatial periodicity of the field. This periodicity is determined by the momentum transferred along the beam trajectory ($\hbar k_{\uvect{v}}=\hbar \omega/v$) since the corresponding wavelength is $\lambda_{\uvect{v}}=2\pi/k_{\uvect{v}}$. Thus, the higher the energy of the polariton (the velocity of the electron) is, the smaller (bigger) $\lambda_{\uvect{v}}$ will be. The wake patterns formed by the field distribution are clearly asymmetric with respect to the beam trajectory. We observe (Fig. \ref{fig6}c) that the wake fields exhibit largest asymmetry as $\alpha$ is increased from $25^{\circ}$ (Fig. \ref{fig6}c, left panel) to $45^{\circ}$ (Fig. \ref{fig6}c, right panel). This is a direct consequence on how the electron transfers different momentum components, $\hbar k_y$ and $\hbar k_z$, to the polaritonic excitation (see Figs. \ref{fig6}b). One can notice from Fig. \ref{fig6}b that as $\alpha$ is increased, $k^{(1)}_z\approx 0$ and $\vect{k}^{(2)}$ tends to the asymptote of the red hyperbola. Therefore, for large angles $\alpha$ the polaritonic wave will propagate relative to the beam trajectory with a phase velocity close to zero on one side of the beam trajectory and with a constant phase velocity on the other side of the beam trajectory. These findings explain the absence of the wavefronts in Fig. \ref{fig6}c for $\alpha=45^{\circ}$ at the left side of the electron beam. It is worth noting that Fig. \ref{fig6}c corresponds to the propagation of the polaritonic wave in the $yz$-plane. However, for other propagation directions, the field distributions will be different.

In Figs. \ref{fig6}d-f we show the same analysis (electron beam trajectory tilted an angle $\alpha$ with respect to the h-BN optical axis) for a representative energy within the lower Reststrahlen band (100\,meV). Importantly, for this case the projections onto the \textit{y}-axis of the phase velocities $\vect{v}^{(1)}_{p}$ and $\vect{v}^{(2)}_{p}$ are antiparallel (negative) to the \textit{y}-component of the Poynting vectors $\vect{S}^{(1)}$ and $\vect{S}^{(2)}$. This yields an asymmetric wave propagating with negative phase velocity (Fig. \ref{fig6}f). 

Additionally, the electron velocity $v$ allows to control the momentum transfer by the fast electron to the phonon polaritons (Eq. \myref{2.g.1}). Indeed, one can obtain the relationship between $v$ and the excitation of the asymmetric wake patterns by analyzing the wake angles $\theta^{(1)}_{\text{w}}=90^{\circ}-
\theta^{(1)}_{\vect{k}}$ and $\theta^{(2)}_{\text{w}}=90^{\circ}-
\theta^{(2)}_{\vect{k}}$ (Figs. \ref{fig6}c,f). We refer the reader to appendix \ref{appendD} where we derived this relationship. For completeness, we show in appendix \ref{appendE} the electron energy losses experienced by a fast electron traveling through h-BN in tilted trajectories with respect to the h-BN optical axis.

We have found that the excitation of the polaritonic wave is highly dependent on the orientation of the electron beam trajectory with respect to the h-BN crystallographic arrangement. Thus, while the speed of the electron serves as a means to excite the polaritonic wave or not, the orientation of the electron beam trajectory can serve to control the direction of the polaritonic excitation.

%
\section{Excitation of Dyakonov surface phonon polaritons in h-BN by a localized beam of fast electrons\label{sec3}}

\begin{figure}[!h]
\begin{center}
\includegraphics[scale=0.67]{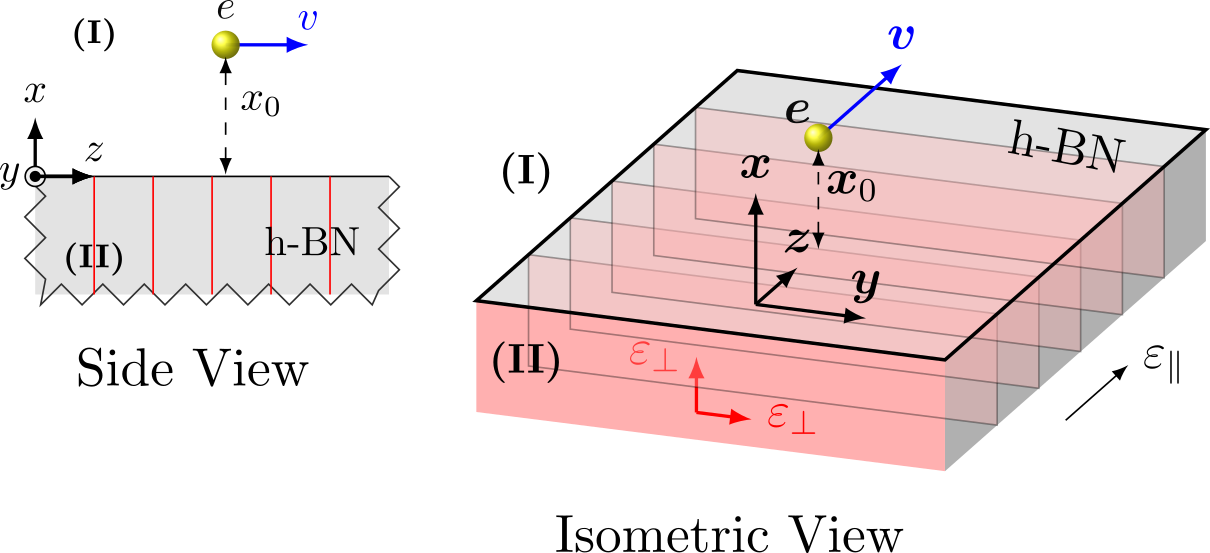}
\caption{Schematics of the probing  electron traveling with velocity $v$ at a distance $x_0$ parallel to a h-BN surface. The optical axis of the h-BN crystal lattice is parallel to the h-BN surface. Label I refers to vacuum, while label II refers to h-BN.}
\label{fig7}
\end{center}
\end{figure}

We next study the EELS signal when the electron beam is traveling above an h-BN semi-infinite surface. The interface between vacuum and h-BN lies on the \textit{yz}-plane, as depicted in Fig. \ref{fig7}, with the \textit{y}-axis in the direction of $\varepsilon_{\bot}$ and the \textit{z}-axis in the direction of the h-BN optical axis. The electron travels in vacuum at a distance $x_0$ from the surface (we will refer to this distance as the impact parameter) with velocity $v$ parallel to the optical axis of the h-BN. A schematic representation of the probing electron-surface system is shown in Fig. \ref{fig7}.

Surfaces of uniaxial materials with optical axis parallel to the surface support a specific kind of surface waves, the so-called Dyakonov waves \cite{Dyakonov1988,Torner2008}. When either $\varepsilon_{\bot}$ or $\varepsilon_{\parallel}$ is negative (as in the case of the Reststrahlen bands in h-BN), surface polaritons called Dyankonov surface polaritons \cite{Talebi2019} can propagate along the surface. Recently, Dyakonov surface phonon polaritons have been observed by scattering-type scanning near-field optical microscopy (s-SNOM) at the edges of h-BN flakes \cite{Li2017,Hillenbrand2017} as well as by STEM-EELS \cite{Konecna2017,Talebi2016}. In the latter experiments, the probing electrons were passing outside the flake edge in a perpendicular trajectory. However, the excitation and detection of Dyakonov surface phonon polaritons with an electron beam parallel to an extended surface has not been described yet.

In the following, we first describe the Dyakonov surface phonon polariton modes that exist at the interface between h-BN and vacuum. We then show that a localized beam of fast electron can couple to these polaritons and we analyse the corresponding EEL spectra and their polariton wake patterns. Importantly, we find that surface Dyakonov phonon polaritons are excited only in the upper Reststrahlen band. Therefore, our analysis and calculations are restricted to this energy range.
%
%
%
%
\subsection{Surfaces modes in h-BN}

According to Dyakonov's theory \cite{Dyakonov1988}, the interface described in Fig. \ref{fig7} supports electromagnetic waves that propagate along it and their associated electromagnetic fields decay exponentially perpendicular to the interface \cite{Dyakonov1988,Torner2008,Cojocaru2014,Gonzalo2019}. These surface waves can be expressed as a linear superposition of the four following modes propagating along the interface: (i) a transverse electric (TE) mode, (ii) a transverse magnetic (TM) mode  [the corresponding  fields decay into the vacuum, upper half space in Fig. \ref{fig7} labeled I], (iii) an ordinary mode, and (iv) an extraordinary mode [the corresponding fields decay exponentially into h-BN, lower half space in Fig. \ref{fig7} labeled II]. Following this scheme the electric field in each media can be written as,
\begin{subequations}
\label{3.a}
\begin{align}
\vect{E}_{\text{I}}(x>0,y,z)&=(\vect{A}_{\text{TE}} + \vect{A}_{\text{TM}}) e^{-\kappa_{\text{I}} x} e^{i (k_y y +k_z z)},\\
\vect{E}_{\text{II}}(x<0,y,z)&=(\vect{A}_{\text{o}} e^{\kappa^{o}_{\text{II}} x} +  \vect{A}_{\text{e}} e^{\kappa^{e}_{\text{II}} x} )e^{i (k_y y +k_z z)},
\end{align} 
\end{subequations}
where harmonic dependency in time has been assumed, and $\vect{A}_{\text{TE}},\vect{A}_{\text{TM}},\vect{A}_{\text{o}},\vect{A}_{\text{e}}$ are the amplitudes of each mode. The wavevector of each mode is given by
\begin{subequations}
\label{3.b}
\begin{align}
\vect{k}_d&=(i\kappa_{\text{I}},k_y,k_z) \hspace{2mm} \text{TE,\, TM},\\
\vect{k}_o&=(-i\kappa^{o}_{\text{II}},k_y,k_z) \hspace{2mm} \text{ordinary},\\
\vect{k}_e&=(-i\kappa^{e}_{\text{II}},k_y,k_z) \hspace{2mm} \text{extraordinary},
\end{align} 
\end{subequations}
where $\kappa_{\text{I}},\kappa^{o}_{\text{II}},\kappa^{e}_{\text{II}}>0$ and $k_y,k_z \in \mathbb{C}$ need to fulfill the following conditions
\begin{subequations}
\label{3.c}
\begin{align}
\label{3.c.a}
&\kappa^2_{\text{I}}=k_y^2+k_z^2-(\omega/c)^2  \hspace{2mm} \text{vacuum},\\
\label{3.c.b}
&(\kappa^{o}_{\text{II}})^2=k_y^2+k_z^2- \varepsilon_{\bot} (\omega/c)^2 \hspace{2mm} \text{ordinary},\\
\label{3.c.c}
&(\kappa^{e}_{\text{II}})^2=k_y^2+\frac{\varepsilon_{\parallel}}{\varepsilon_{\bot}}k_z^2- \varepsilon_{\parallel} (\omega/c)^2 \hspace{2mm} \text{extraordinary}.
\end{align}
\end{subequations}
Applying boundary conditions imposed by Maxwell's equations at the interface between vacuum and h-BN, one obtains the following relationship \cite{Dyakonov1988,Zhu2016,Gonzalo2019}   
\begin{equation}
\label{3.d}
(\kappa_{\text{I}}+ \kappa^{e}_{\text{II}})(\kappa_{\text{I}} + \kappa^{o}_{\text{II}})(\kappa_{\text{I}} + \varepsilon_{\bot} \kappa^{e}_{\text{II}}) =(\omega/c)^2 (\varepsilon_{\parallel}-1)(1-\varepsilon_{\bot})\kappa_{\text{I}}, 
\end{equation}
which together with the set of Eqs. \myref{3.c.a}-\myref{3.c.c} determines the in-plane wavevector $(k_y,k_z)$ of the Dyakonov waves. 

It is worth noting that Dyakonov's original work \cite{Dyakonov1988} was derived for positive $\varepsilon_{\bot}$ and $\varepsilon_{\parallel}$. However Eq. \myref{3.d} is still valid when $\varepsilon_{\bot}<0$ and $\varepsilon_{\parallel}>0$ \cite{Cojocaru2014,Zhu2016}.  Since negative values in the real part of the dielectric components support the excitation of polaritonic states, Dyakonov surface waves sustained in h-BN in the mid-infrared region are thus called Dyakonov surface phonon polaritons.

\begin{figure}[!ht]
\begin{center}
\includegraphics[scale=0.85]{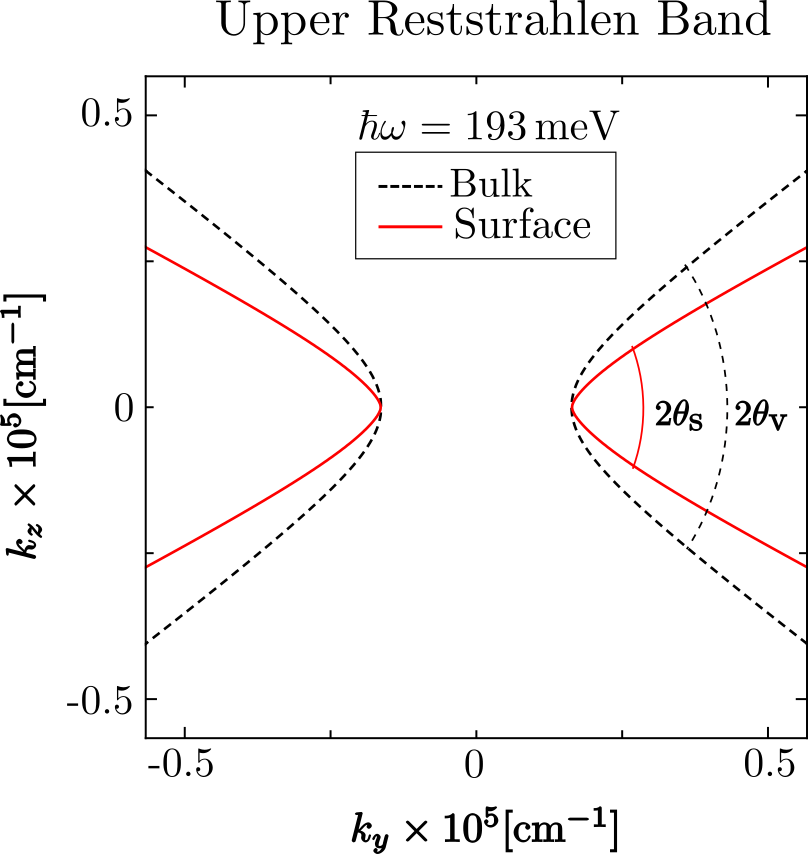}
\caption{The red solid hyperbola represents the isofrequency curve obtained with Eqs. \myref{3.c.a}-\myref{3.c.c} and \myref{3.d} for the surface phonon polariton. While the black dashed hyperbola represents the isofrequency curve obtained using Eq. \myref{2.c} (setting $k_x=0$) for the bulk phonon polariton. Both curves are calculated for a representative energy in the upper Reststrahlen band, 193\,meV.}
\label{fig8}
\end{center}
\end{figure}

In Fig. \ref{fig8} we plot the isofrequency contour (red curve) of the h-BN surface polariton for an energy within the upper Reststrahlen band (193\,meV), obtained from Eqs. \myref{3.c.a}-\myref{3.c.c} and \myref{3.d}. For comparison, we show a cut ($k_yk_z$-plane) of the isofrequency surface of the hyperbolic volume polariton (black dashed line obtained from Eq. \myref{2.c}). We find that the isofrequency curve of the surface polariton is a hyperbola, similar to that of the volume polariton particularly for small momenta. At large momenta, on the other hand, the opening angle of the isofrequency contour of the surface polariton, $\theta_{\text{s}}$, is smaller than that of the volume polariton $\theta_{\text{v}}$, demonstrating that the dispersion of Dyakonov phonon polaritons is different to the one obtained for the bulk hyperbolic phonon polaritons.
%
%

%
%
\subsection{Electron energy loss probability}

The excitation of Dyakonov surface phonon polaritons by fast electron beams can be revealed by electron energy loss spectra. In the following we describe the strategy to obtain the momentum-resolved loss probability, $P^{\text{surf}}(k_y;\omega)$, and the EEL probability, $\Gamma^{\text{surf}}(\omega)$, when the probing electron travels above the h-BN surface (see Fig. \ref{fig7}).

To calculate $\Gamma^{\text{surf}}(\omega)$, following Eq. \myref{2.f}, one needs to obtain the induced electric field, $\vect{E}^{\text{ind}}(\vect{r}_e;\omega)$, along the electron beam trajectory. To that end we obtain $\vect{E}^{\text{ind}}(\vect{r};\omega)$ by solving Maxwell's equations in the presence of vacuum-h-BN interface, assuming that the electron travels in vacuum with constant velocity $v$ and impact parameter $x_0$ (Fig. \ref{fig7}). Considering the boundary conditions at the interfaces ($x=0$), one finds the induced electric field in vacuum (region I in Fig. \ref{fig7}):
\begin{equation}
\label{3.e}
\vect{E}^{\text{ind}}_{\text{I}} (x,k_y,k_z;\omega)=(b_{\text{I}},d_{\text{I}},g_{\text{I}})\,\tilde{\rho}\,e^{-\kappa_{\text{I}} x},
\end{equation} 
with $b_{\text{I}},d_{\text{I}},g_{\text{I}}$ being the coefficients involving the dielectric functions at both sides of the interface and $\tilde{\rho}=-2\pi e \delta(\omega-k_z v) e^{-\kappa_{\text{I}} x_0}/\varepsilon_0$.  We refer to appendices \ref{appendF} and \ref{appendG} for a detailed derivation of the total and induced electric fields at each half space (vacuum and h-BN).

By Fourier transforming $\vect{E}^{\text{ind}}_{\text{I}}(x,k_y,k_z;\omega) \mapsto \vect{E}^{\text{ind}}_{\text{I}}(\vect{r};\omega)$ in Eq. \myref{3.e} and inserting its value into Eq. \myref{2.f}, we find that $\Gamma^{\text{surf}}(\omega)$ can be written as
\begin{align}
\label{3.f}
\Gamma^{\text{surf}}(\omega)= \int_0^{k^c_y} \text{d}k_y \, P^{\text{surf}} (k_y;\omega),
\end{align}
where
\begin{equation}
\label{3.g}
P^{\text{surf}}(k_y;\omega) =  - \frac{e^2}{\pi^2 \epsilon_0 \hbar \omega v}   \text{Re} \left[g_{\text{I}} \,  e^{-2 \kappa_{\text{I}} x_0} \right]\Big\lvert_{k_z=\omega/v},
\end{equation}
is the probability that the electron transfers a transverse momentum $\hbar k_y$ (\textit{y}-component of the momentum) upon loosing energy $\hbar\omega$. Notice that the \textit{z}-component of the wavevector in $P^{\text{surf}}(k_y;\omega)$ is fixed by $k_z=\omega/v$, implying that the electron still transfers a parallel momentum equal to $\hbar \omega/v$.  The integration of Eq. \myref{3.f} is performed up to the  cutoff value $k^c_y$, which is determined by the aperture of the EELS detector.

As we discussed in section \ref{sec2}, the spectrum of the momentum-resolved loss probability and the EEL probability provides information on the properties of the excited modes in the anisotropic medium. We thus show in the following the relationship between these two quantities and the excitation of Dyakonov surface phonon polaritons.
%
%
%
\subsection{Excitation of surface phonon polaritons}
%
%
\begin{figure*}[!ht]
\begin{center}
\includegraphics[scale=0.83]{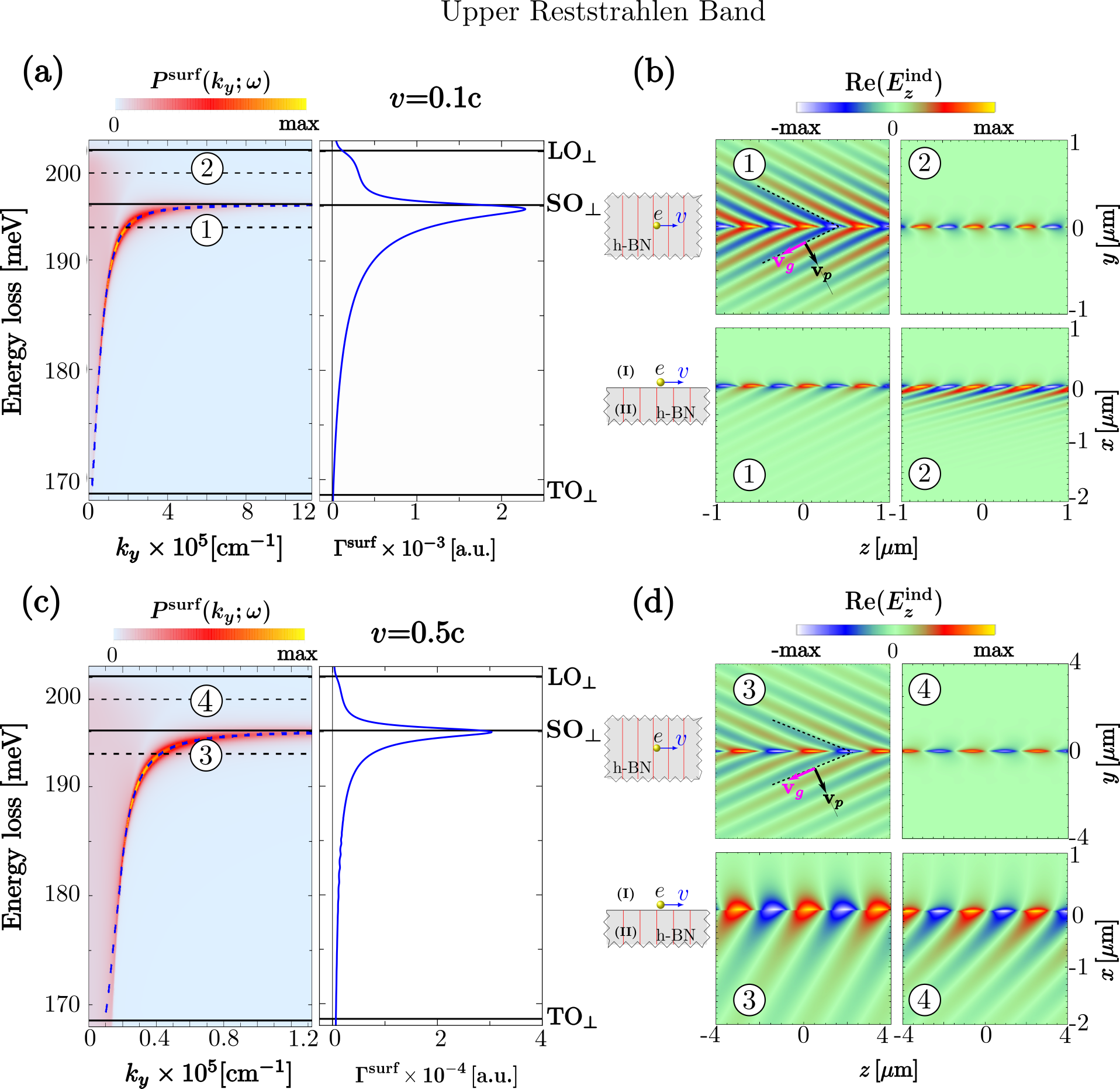}
\caption{The left panel in (a) displays the momentum-resolved loss probability $P^{\text{surf}}(k_y;\omega)$ normalized to the maximum value (3\,a.u.) in the vicinity of the upper Reststrahlen band for $x_0=10\,\text{nm}$ and $v=0.1c$. The right panel in (a) shows the EEL probability $\Gamma^{\text{surf}}(\omega)$ obtained by integrating $P^{\text{surf}}(k_y;\omega)$ over $k_y$ up to $k^c_{y}=0.09\,\mathring{\text{A}}^{-1}$. (c) same as in (a) but considering $v=0.5c$. For this case the maximum value of the momentum-resolved loss probability is 1\,a.u. The color maps in (b) and (d) show the real part of the \textit{z}-component of the induced electric field for the energies: 193 (marked 1, 3) and 198\,meV (marked 2, 4). The top panels in (b) and (d) correspond to the in-plane views ($yz$ plane) of the induced field, while the bottom panels correspond to the out-of-plane views ($xz$ plane).  The field plots are normalized with respect to the maximum value in each case. For the top panels: (b.1) $1\times10^{-4}\,\text{a.u.}$, (b.2) $7.5\times10^{-5}\,\text{a.u.}$, (d.3) $7.5\times10^{-6}\,\text{a.u.}\,$  and (d.4) $5\times10^{-6}\,\text{a.u.}$ For the bottom panels: (b.1) $4\times10^{-5}\,\text{a.u.}$, (b.2) $2\times10^{-5}\,\text{a.u.}$ and (d) $1.5\times10^{-6}\,\text{a.u.}$}
\label{fig9}
\end{center}
\end{figure*}

As pointed out above, the parallel momentum $\hbar k_z$ transferred by the fast electron to the phonon polaritons is determined by the relation $k_z=\omega/v$. Similarly to the bulk analysis of section \ref{sec2}, this relationship represents a horizontal line in the $k_yk_z$ representation of Fig. \ref{fig8}. Thus, the transferred momentum can be determined by the crossing between this horizontal line ($k_z=\omega/v$) and the isofrequency hyperbolas obtained from Eqs. \myref{3.c.a}-\myref{3.c.c} and \myref{3.d}. From the latter equations one can obtain the relationship between the \textit{y}-component of the polariton wavevector, $k_y$, and $\hbar\omega$, which is shown in the left panel of Fig. \ref{fig9}(a) (dashed blue curve). We also plot the momentum-resolved loss probability $P^{\text{surf}}(k_y;\omega)$ for energies around the upper Reststrahlen band. The probing electron is traveling above the h-BN surface with an impact parameter of 10\,nm and a velocity $v=0.1c$. Some similarities between $P^{\text{surf}}(k_y;\omega)$ and $P^{\text{bulk}}(k_{\bot};\omega)$ (Fig. \ref{fig3}b, left panel) become apparent. For instance, the highest values of $P^{\text{surf}}(k_y;\omega)$ (red and yellow colors in Fig. \ref{fig9}a) coincide perfectly with the blue dashed curve, demonstrating that the electron energy losses in the upper band are caused mainly due to the excitation of hyperbolic phonon polaritons. However, by comparing Figs. \ref{fig3}b and \ref{fig9}a we recognize that the asymptotic behavior (at large momenta) of $P^{\text{surf}}(k_y;\omega)$ occurs at a lower energy compared to the asymptotic behavior of $P^{\text{bulk}}(k_{\bot};\omega)$. While $P^{\text{bulk}}(k_{\bot};\omega)$ tends to the $\text{LO}_{\bot}$ phonon energy, $P^{\text{surf}}(k_y;\omega)$ tends to the surface optical ($\text{SO}_{\bot}$) phonon energy given by the condition $\varepsilon_{\bot}(\omega_{\text{SO}_{\bot}})=-1$ (derived from Eqs. \myref{3.c.a}-\myref{3.c.c}  and \myref{3.d} for large momenta). Importantly, the latter is a fingerprint of the excitation of surface polariton modes. In our case (electron traveling in vacuum above the h-BN surface) these surface modes correspond to Dyakonov surface phonon polaritons. We confirm this by integrating $P^{\text{surf}}(k_y;\omega)$ over $k_y$ up to a cutoff momentum $\hbar k^c_y$, which yields the EEL probability $\Gamma^{\text{surf}}(\omega)$ (right panel of Fig. \ref{fig9}b). A clear peak can be observed at the $\text{SO}_{\bot}$ phonon energy. This loss peak is slightly asymmetric with a broader tail for lower energies in the Reststrahlen band compared to that for larger energies in the band. Notice that for energies above $\text{SO}_{\bot}$ the loss spectrum displays a shoulder arising from background losses present in the entire upper band at small momentum (red blurred area for small momentum in the left panel of Fig. \ref{fig9}a). 

The excitation of Dyakonov surface phonon polaritons (within the upper Reststrahlen band) by the probing electron can be observed in real space in Fig. \ref{fig9}b, where we show the real part of the \textit{z}-component of the induced electric field at energies 193\,meV (marked as 1) and 198\,meV (marked as 2). The top panels correspond to the evaluation of $\text{Re}(E^{\text{ind}}_z(\vect{r};\omega))$ in the $yz$-plane (in-plane at the interface), and the bottom panels to the evaluation in the $xz$-plane (lateral view, containing the electron trajectory). One can recognize from the in-plane views (Figs. \ref{fig9}b marked as 1) the formation of wake patterns and the oscillatory behavior of the induced field in the \textit{z}-direction. Similarly to the field distribution shown in Fig. \ref{fig3}c, the spatial periodicity along the \textit{z}-direction is connected with the parallel wavevector component $k_z=\omega/v$, since $\lambda_z=2\pi/k_z$. Moreover, the wake wavefronts show interesting propagation patterns both in the transverse direction from the beam trajectory as well as into the h-BN. 

In the top panel of Fig. \ref{fig9}b (image labeled as 1), the wavefronts along the \textit{y}-direction propagate with positive phase and group velocities relative to the Poynting vector. Indeed, we find that the dashed blue curve in Fig. \ref{fig9}a has a positive slope ($\text{d}\omega/\text{d} k_y>0$), indicating that the projections onto the \textit{y}-axis of the group and phase velocities are parallel (positive). We also notice that Dyakonov surface phonon polaritons are confined to the interface with penetration of the field into the h-BN interface (Fig. \ref{fig9}b, bottom image labeled as 1). For energies larger than that of the $\text{SO}_{\bot}$ phonon, Dyakonov surface phonon polaritons are not excited (Fig. \ref{fig9}b, image labeled as 2). Thus, the induced field distributions for those energies correspond to the reflection of the electron electromagnetic field at the h-BN surface (Fig. \ref{fig9}b, top image labeled as 2). We can also notice that the field penetrates into the h-BN (bottom panel 2 of Fig. \ref{fig9}b), which is connected with the presence of the red blurred region corresponding to the losses appearing for lower momenta in Fig. \ref{fig9}a (left panel).

When the velocity of the probing electron is increased up to 50\,\% the speed of light, the momentum parallel to the beam trajectory, $\hbar k_z$, is reduced and so does the $k_y$ component of the Dyakonov surface phonon polariton. By calculating the momentum-resolved loss probability $P^{\text{surf}}(k_y;\omega)$ (left panel of Fig. \ref{fig9}c) and the EEL probability $\Gamma^{\text{surf}}(\omega)$ one obtains a similar behavior as in Fig. \ref{fig9}a for $v=0.1c$, except for a one order of magnitude reduction of both $k_y$ and the value of the loss probability. 

The differences in the properties of the Dyakonov surface phonon polaritons launched by the fast electron beam can be observed in Fig. \ref{fig9}d, where we show the real part of the \textit{z}-component of the induced electric field for $v=0.5c$ at energies 193\,meV and 200\,meV. Notice that the spatial periodicity $\lambda_z$ of the polariton is longer in this situation compared to that in Fig. \ref{fig9}b as a result of the increased electron velocity. Also, the penetration of the field into the h-BN medium is larger compared to that  in Fig. \ref{fig9}b. This increase in the penetration depth can be attributed to the increase of the background losses present in the entire upper band (blurred red are in the left panel of Fig. \ref{fig9}c).
 
For completeness and similar to the analysis presented above, we study in appendix \ref{appendH} the excitation of phonon polaritons in the lower Reststrahlen band by a fast electron traveling in aloof trajectory. In the appendix we show the momentum-resolved loss probability ($P^{\text{surf}}(k_y;\omega)$), the EEL probability ($\Gamma^{\text{surf}}(\omega)$) and the wake patterns for this energy range.
%
%
\section{Remote excitation of bulk phonon polaritons\label{sec4}}
We have shown in Figs. \ref{fig9}b and \ref{fig9}d that the electric field penetrates into the bulk of the h-BN semi-infinite surface, which is surprising, as one does not expect the excitation of volume modes in isotropic materials for electron beam trajectories outside the material. By comparing the angles of the wake patterns, we demonstrate that indeed volume modes are excited in h-BN with external beam trajectories.
%
%
\begin{figure}[!ht]
\begin{center}
\includegraphics[scale=0.7]{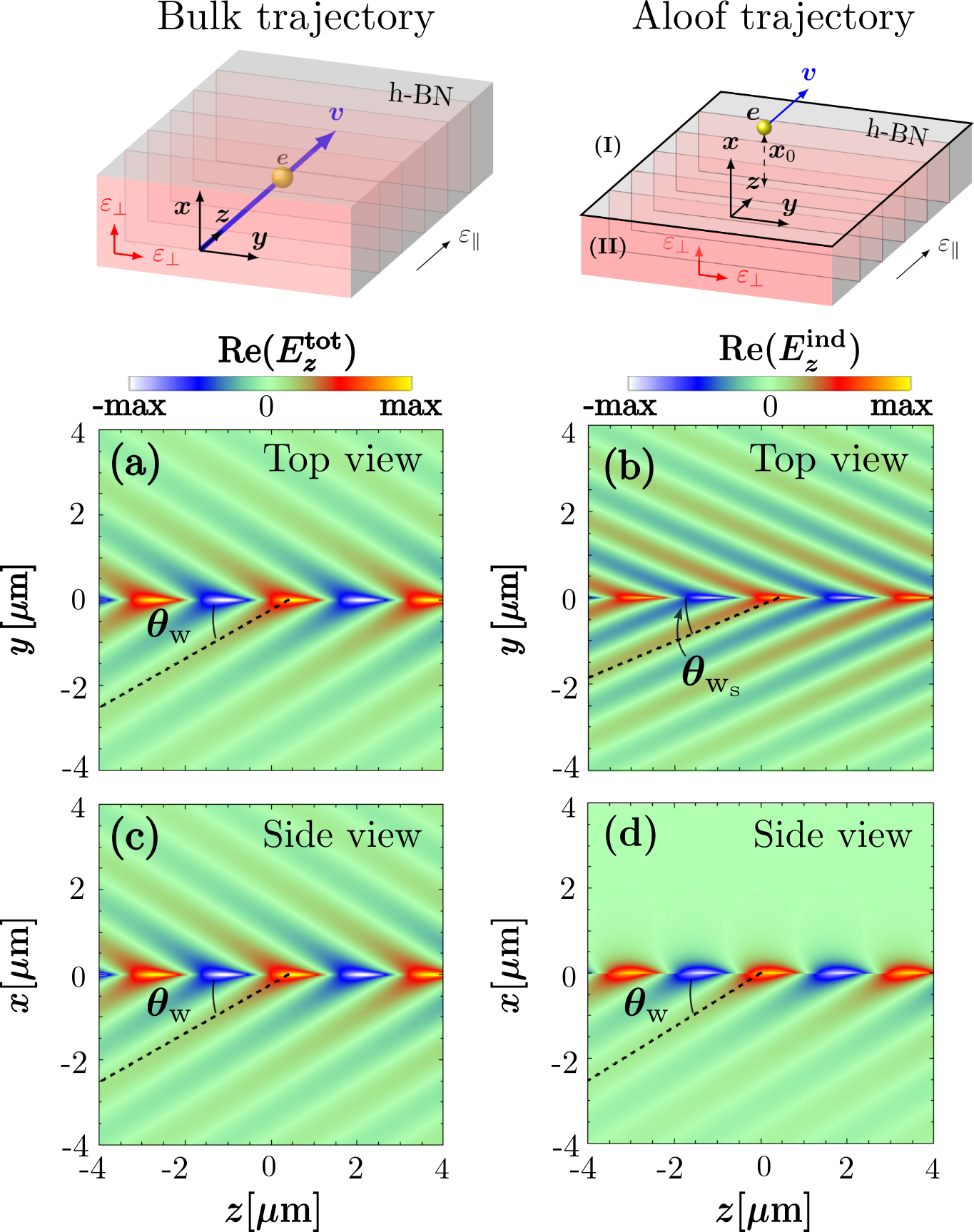}
\caption{(a) Real part of the \textit{z}-component of the total electric field in the \textit{yz}-plane produced by a fast electron traveling through h-BN parallel to its optical axis. (c) shows $\text{Re}(E^{\text{tot}}_{z})$ evaluated in the \textit{xz}-plane and $\theta_{\text{w}}$ is the angle between the \textit{z}-axis and the wake patterns formed by the bulk polariton. (b) shows the real part of the \textit{z}-component of the induced electric field produced by a fast electron traveling in vacuum 10\,nm above a semi-infinite h-BN surface. (d) shows $\text{Re}(E^{\text{ind}}_{z})$ evaluated in the \textit{xz}-plane and $\theta_{\text{w}_{\text{s}}}$ is the angle between the \textit{z}-axis and the wake patterns formed by the Dyakonov surface phonon polariton. We used for the calculation of the fields an electron velocity equal to $v=0.5c$ at energy $\hbar\omega=193\,\text{meV}$. The field plots are normalized with respect to the maximum value in each case: (a) $5\times10^{-7}\,\text{a.u.}$, (b) $7.5\times10^{-6}\,\text{a.u.}$, (c) $5\times10^{-7}\,\text{a.u.}\,$  and (d) $1.5\times10^{-6}\,\text{a.u.}$ The insets above (a) and (b) illustrate the geometry under consideration for each case.}
\label{fig10}
\end{center}
\end{figure}

We first calculated the angle $\theta_{\text{w}}$ of the wake wavefronts produced by the fast electron traveling through bulk h-BN with $v=0.5c$ at $\hbar\omega=193\,\text{meV}$ (Figs. \ref{fig10}a and \ref{fig10}c), obtaining a value of $\theta_{\text{w}}=32.35^{\circ}$. We compare $\theta_{\text{w}}$ with the angles of the wake wavefronts produced by the fast electron traveling in an aloof trajectory 10\,nm above the h-BN surface (Fig. \ref{fig10}b and \ref{fig10}d). From this comparison we find that: (i) the angle $\theta_{\text{w}_{\text{s}}}=24.67^{\circ}$ of the wake pattern at the h-BN surface (Fig. \ref{fig10}b) is different from $\theta_{\text{w}}$, and (ii) the angle of the wake pattern excited inside the h-BN is the same as $\theta_{\text{w}}$ (Fig. \ref{fig10}d).  This implies that volume modes are excited by the fast electron traveling in trajectories outside the anisotropic medium.

Importantly, these findings open the possibility of remotely exciting volume phonon polaritons. In contrast to isotropic materials, where an aloof electron beam only couples to surface modes, for anisotropic materials the energy and momentum matching between the electron and the polaritons allows for launching of bulk excitations.

%
\section{Summary\label{sec5}}

We have thoroughly analyzed the excitation of optical phonon polaritons in hexagonal boron nitride by focused electron beams for two relevant situations: when the electron travels through the h-BN bulk and when it travels in vacuum above a semi-infinite h-BN surface. For the bulk situation, we have observed that the electron couples to volume phonon polaritons. We demonstrated that the excitation of these polaritonic modes is strongly dependent on the electron velocity and on the angle between the optical axis of h-BN and the trajectory of the electron beam.  Furthermore, we have shown that Dyakonov surface phonon polaritons can be excited by a fast electron traveling above the h-BN surface. Interestingly, aloof electron beams are capable of exciting volume polaritons in the h-BN.

By a detailed mode analysis, we showed that the electron beam transfers a specific momentum to the modes. This momentum transfer determines the properties of the excited phonon polaritons, and thus controls their phase and group velocities, as well as their propagation direction. Importantly, we found that the propagation of the polaritonic waves is highly asymmetric with respect to the electron beam trajectory when the trajectory sustains an angle relative to the h-BN optical axis.

Our findings may offer a way to steer and control the propagation of the polaritonic waves excited in hyperbolic materials. Although we studied the specific material h-BN, our findings can be generalized and can serve as an guide for the correct interpretation of the different excited modes and loss channels encountered in EELS experiments of uniaxial materials.  

%
\section{Acknowledgements\label{sec6}}

C.M.-E. thanks \'{A}lvaro Nodar for his help in the parallelization of the programming codes. J. A. and R. H. acknowledges Grant No. IT1164-19 for research groups of the Basque University system from the Department of Education of the Basque Government. R.H. further acknowledges financial support from the Spanish Ministry of Science, Innovation and Universities (national project RTI2018-094830-B-100 and the project MDM-2016-0618 of the Marie de Maeztu Units of Excellence Program).
%
%
\appendix 

%
\section{h-BN dielectric function\label{appendA}}
The two components of the h-BN dielectric function can be described by a Drude-Lorentz model as \cite{Caldwell2014}
\begin{equation}
\label{a.1}
\varepsilon(\omega) = \varepsilon_{\infty} \left(1 + \frac{\omega^2_{\text{LO}} - \omega^2_{\text{TO}}}{\omega^2_{\text{TO}} - \omega^2 - i \omega \gamma}  \right), 
\end{equation}
with $\hbar\omega_{\text{LO}}$, $\hbar\omega_{\text{TO}}$ the phonon LO, TO energies, respectively, $\varepsilon_{\infty}$ is the high-frequency dielectric permittivity and $\gamma$ is the damping constant. The values used for each constant are presented in Table I.
\begin{table}[h]
\label{table1}
\caption{Parameters used for the in-plane and out-of-plane dielectric components within the Drude-Lorentz model taken from reference \citenum{Hillenbrand2017}.}
\begin{ruledtabular}
\begin{tabular}{ccc}
&In-plane ($\varepsilon_{\bot}$) & \mbox{Out-of-plane ($\varepsilon_{\parallel}$)} \\
\hline
$\varepsilon_{\infty}$& 4.90 & \mbox{2.95} \\
$\hbar\omega_{\text{TO}}$ & 168.6 meV & 94.2 meV \\
$\hbar\omega_{\text{LO}}$ & 200.1 meV & 102.3 meV \\
$\gamma$ & 0.87 meV & 0.25 meV \\
\end{tabular}
\end{ruledtabular}
\end{table}

%
\section{Green's tensor decomposition in an anisotropic medium \label{appendB}}
In this work we use the following Fourier transform convention
\begin{equation}
\label{b.1}
\hat{\vect{F}}(\vect{k};\omega)= \int_{-\infty}^{\infty}  \text{d}t\,  \int_V \text{d}^3{\vect{r}} \, \vect{F} (\vect{r};t) \, e^{i(\omega t - \vect{k} \cdot \vect{r})},
\end{equation}
where $\vect{F}(\vect{r};t)$ is a smooth vector field representing the electric or magnetic fields and $V$ stands for the volume in the euclidean space $\mathbb{R}^3$. Thus, the Green's tensor satisfying the wave equation \cite{Chew1995,DeAbajo2008,Collin,Novotny} 
\begin{equation}
\label{b.2}
\nabla^2 {\tensor{\bf G}(\vect{r};\omega)} + k^2_0 \tensor{\varepsilon} \tensor{\bf G}(\vect{r};\omega)- \nabla[\nabla \cdot \tensor{\bf G}(\vect{r};\omega)]=\tensor{\bf I}\delta(\vect{r}), 
\end{equation}
can be expressed in $k-\omega$ space as follows
\begin{equation}
\label{b.3}
\tensor{\bf G}(\vect{k};\omega)= \left[\vect{k} \otimes \vect{k} -k^2 {\tensor{\bf I}} + k^2_0 {\tensor{\varepsilon}}\right]^{-1}.
\end{equation}
From Eq. \myref{b.3} one can deduce that the inverse of the Green's tensor for an uniaxial medium can be decomposed in the form $\tensor{\bf G}^{-1}= (k^2_0\varepsilon_{\bot} - k^2)\, \tensor{\bf I} + \vect{k} \, \otimes \, \vect{k} + k^2_0 (\varepsilon_{\parallel}-\varepsilon_{\bot})\, \uvect{z}\, \otimes \, \uvect{z}$ where $\varepsilon_{\bot}=\varepsilon_x=\varepsilon_y$ and $\varepsilon_{\parallel}=\varepsilon_z$. This tensor decomposition allows for finding the following closed expression for $\tensor{\bf G}(\vect{k};\omega)$\cite{Eroglu2010}:
\begin{align}
\label{b.4}
\tensor{\bf G}(\vect{k};\omega)&=\frac{1}{k^2_0 \varepsilon_{\parallel} \varepsilon_{\bot} - \vect{k} \cdot \tensor{\varepsilon}\cdot \vect{k}} \left[ \varepsilon_{\parallel} \tensor{\bf I} - (\varepsilon_{\parallel} - \varepsilon_{\bot}) \uvect{z} \otimes \uvect{z}  \right.\\
\nonumber
&\hspace{4mm} \left. - \frac{\vect{k} \otimes \vect{k}}{k^2_0} + \frac{\varepsilon_{\parallel}- \varepsilon_{\bot}}{k^2_0 \varepsilon_{\bot} - k^2} (\vect{k} \times \uvect{z}) \otimes (\vect{k} \times \uvect{z})  \right],
\end{align}
where we used that the inverse of the Green's tensor can be obtained as $\tensor{\bf G}^{-1}=\text{adj}[{\tensor{\bf G}}]/\text{det}[{\tensor{\bf G}}]$, with $\text{adj}[{\tensor{\bf G}}]$ the adjoint of the Green's tensor. 
%
%
\section{Bulk EEL probability for different cutoff values  $k^c_{\bot}$ \label{appendC}}
\begin{figure}[!ht]
\begin{center}
\includegraphics[scale=0.9]{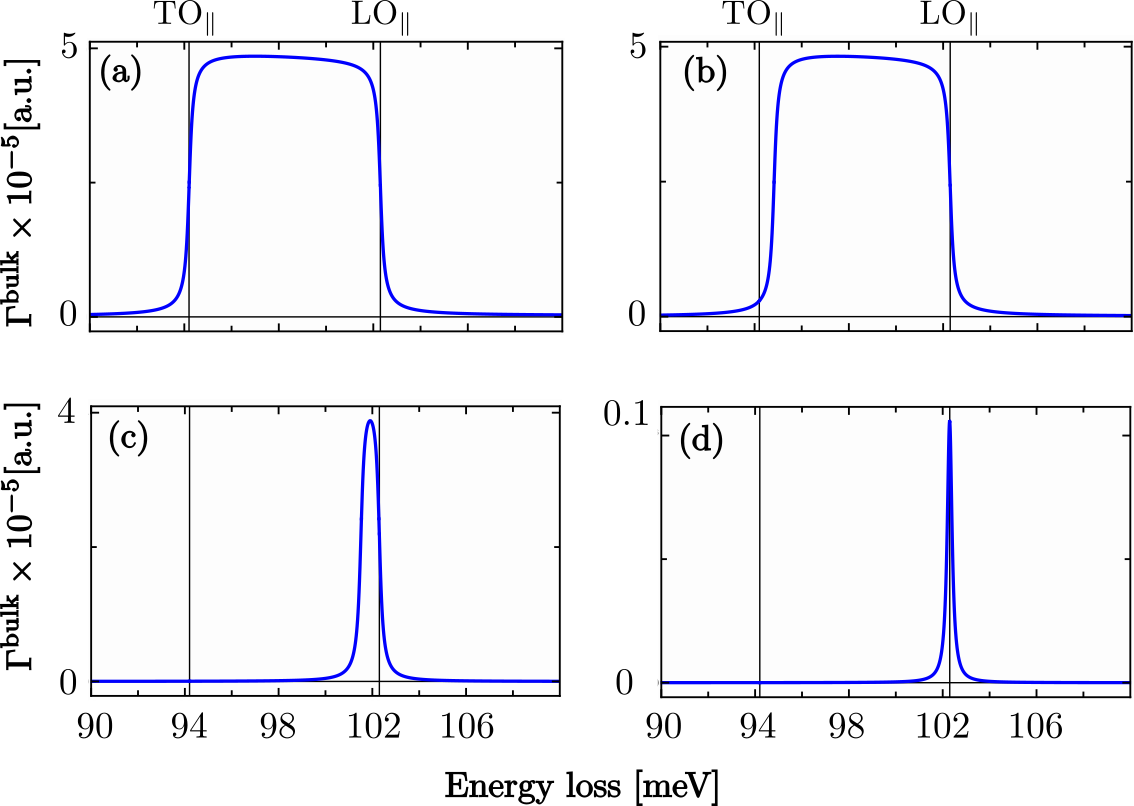}
\caption{Electron energy loss probability, $\Gamma^{\text{bulk}}(\omega)$, for energies around the lower Reststrahlen band for four different $k^c_{\bot}$: (a) $1 \times 10^{-2}\,\mathring{\text{A}}^{-1}$, (b) $1 \times 10^{-3}\,\mathring{\text{A}}^{-1}$, (c) $1 \times 10^{-4}\,\mathring{\text{A}}^{-1}$ and (d) $1 \times 10^{-5}\,\mathring{\text{A}}^{-1}$. The electron travels through h-BN parallel to the optical axis with velocity $v=0.1c$.}
\label{fig11}
\end{center}
\end{figure}

In Fig. \ref{fig11} we show the EEL probability ($\Gamma^{\text{bulk}}(\omega)$, given by Eq.  \myref{2.m}) in the vicinity of the lower Reststrahlen band for different cutoff values $k^c_{\bot}$: (a) $1 \times 10^{-2}\,\mathring{\text{A}}^{-1}$, (b) $1 \times 10^{-3}\,\mathring{\text{A}}^{-1}$, (c) $1 \times 10^{-4}\,\mathring{\text{A}}^{-1}$ and (d) $1 \times 10^{-5}\,\mathring{\text{A}}^{-1}$.  For the calculation of $\Gamma^{\text{bulk}}(\omega)$ we consider $v=0.1c$.

One can observe that for small cutoff momentum the EEL probability of the $\text{LO}_{\parallel}$ phonon energy is better defined. Whereas for large cutoff momenta the sharp peak in Fig. \ref{fig11}d broadens. However, cutoff values of $1\times 10^{-4}\,\mathring{\text{A}}^{-4}$ or $\times 10^{-4}\,\mathring{\text{A}}^{-5}$ are not experimentally feasible.
%
%
%
%
%
\section{Analysis of the asymmetries of bulk polaritonic waves \label{appendD}}
When the electron beam trajectory makes an angle $\alpha$ relative to the h-BN optical axis, the propagation of the phonon polaritons (excited by the fast electron) is highly asymmetric with respect to the beam trajectory. We analyze these asymmetries in the following. 
%
%
\begin{figure*}[!ht]
\begin{center}
\includegraphics[scale=0.68]{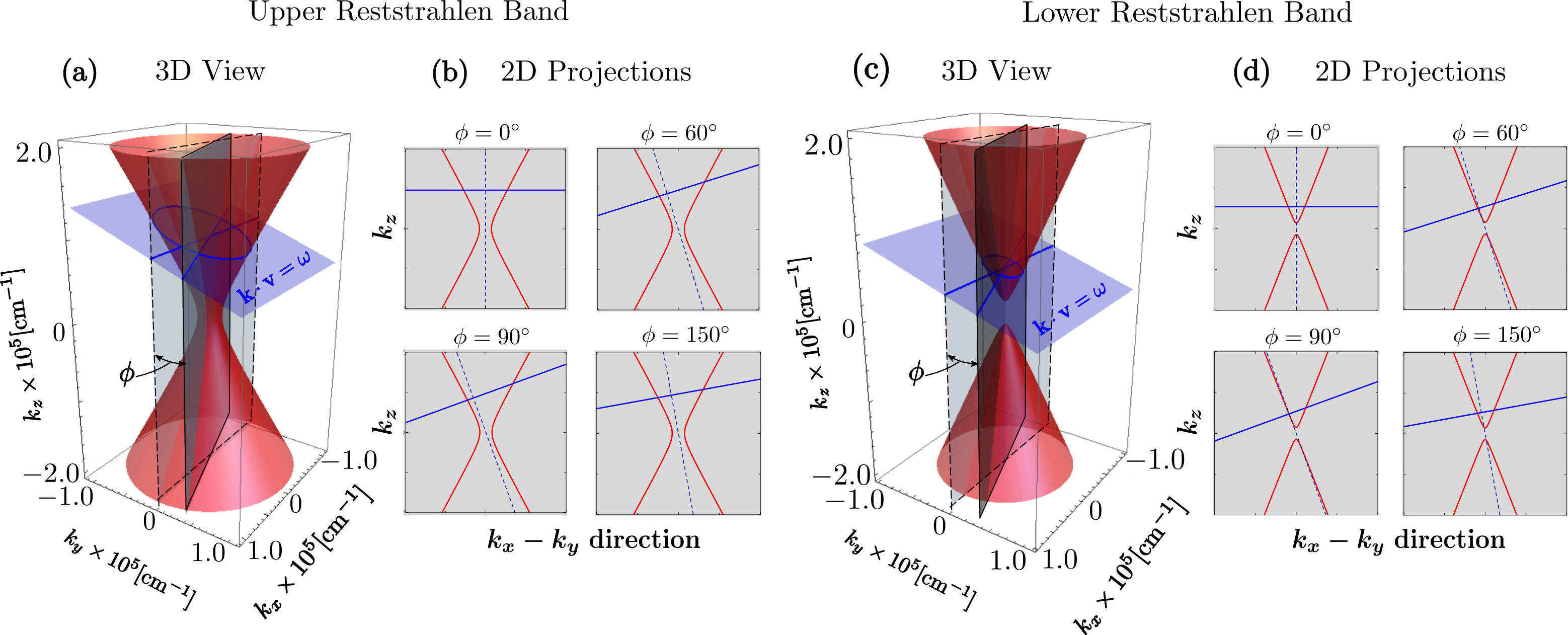}
\caption{(a) Isofrequency surface (red hyperboloid) for a representative energy in the upper Reststrahlen band (180\,meV). The blue inclined plane depicts Eq. \myref{d.1.b} for an electron beam with $v=0.1c$ and trajectory angle of $\alpha=20^{\circ}$. The grey plane represents the different directions set by the azimuthal angle $\phi$. The 2D plots in (b) show the intersection between the red hyperboloid and the blue plane in the four different directions determined by $\phi$: $0^{\circ},60^{\circ},90^{\circ}$ and $150^{\circ}$. The blue dashed lines in the 2D projections depict the trajectory of the electron beam, as viewed along each direction determined by the angle $\phi$. (c) and (d) are the same as (a) and (b) but for a representative energy in the lower Reststrahlen band (100\,meV).}
\label{fig12}
\end{center}
\end{figure*}
%

The propagation of the polaritonic wave is governed by its phase velocity and thus, by the polariton wavevector $\vect{k}(\omega)=(k_x,k_y,k_z)$ which fulfills Eq. \myref {2.c}. When the hyperbolic phonon polaritons are excited by an electron beam, the components of $\vect{k}(\omega)$ have also to fulfill Eq. \myref{2.g.1}, that is, the components of $\vect{k}(\omega)$ can be obtained from the following two expressions 
\begin{subequations}
\label {d.1}
\begin{align}
\label{d.1.a}
&\frac{k^2_x+k^2_y}{\varepsilon_{\parallel}}+\frac{k^2_z}{\varepsilon_{\bot}}=k^2_0,\\
\label{d.1.b}
& k_y \sin \alpha + k_z \cos \alpha = \omega / v,
\end{align}
\end{subequations}
where we assume that the electron velocity is $\vect{v}=v(0,\sin\alpha,\cos\alpha)$. Moreover, if we decompose $\vect{k}(\omega)$ in cylindrical coordinates as $\vect{k}(\omega)=(q\cos\phi,q\sin\phi,k_z)$, with $\phi$ the azimuthal angle of the symmetry axis, and substitute it into Eqs. \myref{d.1.a} and \myref{d.1.b}, we obtain to the following system of equations
\begin{subequations}
\label{d.2}
\begin{align}
\label{d.2.a}
&\frac{q^2}{\varepsilon_{\parallel}}+\frac{k^2_z}{\varepsilon_{\bot}}=k^2_0,\\
\label{d.2.b}
& q \sin \phi \sin \alpha + k_z \cos \alpha = \omega / v,
\end{align}
\end{subequations}
for $q,\phi$ and $k_z$. Notice that the variable $q$ corresponds to $k_{\bot}$ for trajectories parallel to the h-BN optical axis. However, for the oblique trajectory $\hbar \vect{q}=(\hbar k_x,\hbar k_y)$ is no longer orthogonal to the beam trajectory and thus we avoid referring to it as the transverse momentum. One can deduce from Eqs. \myref{d.2.a} and \myref{d.2.b} that the solutions have cylindrical symmetry (symmetric with respect to the \textit{z}-axis) when $\alpha=0^{\circ}$. For cases where $\alpha\neq 0^{\circ}$, this symmetry is broken and the solutions depend on the azimuthal angle $\phi$. We explore this dependency below. 

In Fig. \ref{fig12} we show the intersection between the h-BN isofrequency hyperboloids (red surfaces, Figs. \ref{fig12}a,c) and the plane $\vect{k}\cdot\vect{v}=\omega$ determined by the electron beam trajectory (blue surfaces, Figs. \ref{fig12}a,c). Notice that the direction of the electron beam trajectory is orthogonal to the blue plane $\vect{k}\cdot\vect{v}=\omega$. We analyze an electron beam with velocity $v=0.1c$ and a trajectory angle of $\alpha=20^{\circ}$. Finally we chose  two representative energies,  one in the upper Reststrahlen band at 180\,meV (Fig. \ref{fig12}a) and the other one in the lower Reststrahlen band at 100\,meV (Fig. \ref{fig12}c). The grey 2D plots in Figs. \ref{fig12}b and \ref{fig12}d show the intersection between the red hyperboloid and the blue plane along four different directions determined by the azimuthal angle $\phi$: $0^{\circ},60^{\circ},90^{\circ}$ and $150^{\circ}$. In the 2D projections the blue dashed lines depict the beam trajectory, as viewed from the direction determined by $\phi$. The polariton wavevector along each particular direction can be obtained from the intersection between the blue lines and the red hyperbolas. Importantly, one can recognize from the 2D projections that:
\begin{enumerate}
\item The intersection between the blue line and the red hyperbola is asymmetric with respect to the \textit{z}-axis for $\phi\in (0^{\circ},180^{\circ})$, as we observe in Figs. \ref{fig12}(b) and \ref{fig12}(d) for $\phi=60^{\circ},90^{\circ},150^{\circ}$.
\item The direction of largest asymmetry occurs at $\phi=90^{\circ}$ ($k_yk_z$-plane) and the direction of symmetric propagation occurs at $\phi=0^{\circ}$ ($k_xk_z$-plane).
\item The intersections between the blue lines and the red hyperbolas are also asymmetric (or symmetric) with respect to the electron beam trajectory (blue dashed line). 
\end{enumerate}  

To better understand the asymmetries in the propagation of the polaritonic waves, we focus on the direction of largest asymmetry: $\phi=90^{\circ}$ (equivalently the $k_yk_z$-plane). From Eqs. \myref{d.1.a} and \myref{d.1.b} one can obtain the following two solutions for the polariton wavevector in the $k_yk_z$-plane
\begin{subequations}
\label{d.3}
\begin{align}
\label{d.3.a}
\vect{k}^{(1)}&=\frac{\omega}{v}\left[\frac{\varepsilon_{\parallel} \sin\alpha + \sqrt{\varepsilon_{\parallel} \varepsilon_{\bot}\Delta} \cos \alpha}{\varepsilon_{\bot} \cos^2\alpha + \varepsilon_{\parallel} \sin^2\alpha} \right] \uvect{y} \\
\nonumber
&\hspace{4mm} + \frac{\omega}{v}\left[ \frac{\varepsilon_{\bot} \cos\alpha - \sqrt{\varepsilon_{\parallel} \varepsilon_{\bot}\Delta} \sin \alpha}{\varepsilon_{\bot} \cos^2\alpha + \varepsilon_{\parallel} \sin^2\alpha} \right] \uvect{z}, \\
\label{d.3.b}
\vect{k}^{(2)}&=\frac{\omega}{v}\left[\frac{\varepsilon_{\parallel} \sin\alpha - \sqrt{\varepsilon_{\parallel} \varepsilon_{\bot}\Delta} \cos \alpha}{\varepsilon_{\bot} \cos^2\alpha + \varepsilon_{\parallel} \sin^2\alpha} \right] \uvect{y} \\
\nonumber
&\hspace{4mm} + \frac{\omega}{v}\left[ \frac{\varepsilon_{\bot} \cos\alpha + \sqrt{\varepsilon_{\parallel} \varepsilon_{\bot}\Delta} \sin \alpha}{\varepsilon_{\bot} \cos^2\alpha + \varepsilon_{\parallel} \sin^2\alpha} \right] \uvect{z}.
\end{align}
\end{subequations}
with
\begin{equation}
\label{d.4}
\Delta=\left(\frac{v}{c} \cos \alpha \right)^2 \varepsilon_{\bot} + \left(\frac{v}{c} \sin \alpha \right)^2 \varepsilon_{\parallel} -1.
\end{equation}
From Eqs. \myref{d.3.a} and \myref{d.3.b} one can recognize that $k^{(1)}_z\neq k^{(2)}_z$, showing the asymmetry in the propagation of the polaritonic wave. Moreover, the angles $\theta^{(1)}_{\vect{k}}$ and  $\theta^{(2)}_{\vect{k}}$ defined by $\vect{k}^{(1)}$, $\vect{k}^{(2)}$ vectors with respect to the electron beam trajectory (see Figs. \ref{fig6}b and \ref{fig6}e) satisfy the following relations
\begin{subequations}
\label{d.5}
\begin{align}
\label{d.5.a}
\tan(\theta^{(1)}_{\vect{k}} + \alpha) &= \frac{\varepsilon_{\parallel} \sin\alpha +\sqrt{\varepsilon_{\parallel}\varepsilon_{\bot} \Delta} \cos\alpha}{\varepsilon_{\bot} \cos\alpha - \sqrt{\varepsilon_{\parallel}\varepsilon_{\bot}\Delta}\sin\alpha},\\
%
\label{d.5.b}
\tan(\theta^{(2)}_{\vect{k}}- \alpha) &= \frac{\varepsilon_{\parallel} \sin\alpha - \sqrt{\varepsilon_{\parallel}\varepsilon_{\bot} \Delta} \cos\alpha}{\varepsilon_{\bot} \cos\alpha + \sqrt{\varepsilon_{\parallel}\varepsilon_{\bot}\Delta}\sin\alpha}.
\end{align}
\end{subequations}
When $\alpha=0^{\circ}$, one can deduce from Eqs.  \myref{d.5.a} and \myref{d.5.b} that
\begin{subequations}
\label{d.6}
\begin{align}
\label{d.6.a}
\tan\theta^{(1)}_{\vect {k}}&=\sqrt{\left(\frac{v}{c}\right)^2\varepsilon_{\parallel}-\frac{\varepsilon_{\parallel}}{\varepsilon_{\bot}}},\\
\label{d.6.b}
\tan\theta^{(2)}_{\vect {k}}&=-\sqrt{\left(\frac{v}{c}\right)^2\varepsilon_{\parallel}-\frac{\varepsilon_{\parallel}}{\varepsilon_{\bot}}}.
\end{align}
\end{subequations}
Therefore $\theta^{(1)}_{\vect {k}}=\theta^{(2)}_{\vect {k}}=\theta_{\vect{k}}$ for this particular case of symmetric propagation. Notice that $\theta_{\vect{k}}$ is also preserved in any other azimuthal direction.

We can observe from Eqs. \myref{d.3.a} and \myref{d.3.b} that $\vect{k}^{(1)},\vect{k}^{(2)}$ depend on the electron velocity $v$. This dependency provides information on the condition that the electron velocity needs to satisfy for the electron beam to excite the polaritonic waves. Indeed, by imposing real value solutions to Eqs. \myref{d.5.a} and \myref{d.5.b}, one obtains the following condition on $v$:
\begin{equation}
\label{d.7}
\frac{v^2}{c^2}\left[ \varepsilon^2_{\bot} \varepsilon_{\parallel} \cos^2\alpha + \varepsilon^2_{\parallel} \varepsilon_{\bot} \sin^2\alpha  \right]>\varepsilon_{\bot}\varepsilon_{\parallel}.
\end{equation}
This last relationship results in the following inequality
\begin{equation}
\label{d.8}
\frac{v^2}{c^2} \varepsilon_{\parallel} >\frac{\varepsilon_{\parallel}}{\varepsilon_{\bot}},
\end{equation}
when $\alpha=0^{\circ}$, which coincides exactly with the first inequality in Eq. \myref{2.n} obtained in the main text. As we discuss in section \ref{sec2}E, Eq. \myref{d.8} reveals the condition on the electron velocity for exciting phonon polaritons or emitting Cherenkov radiation.

We show now that we can recover the properties of the excited wave in an isotropic dielectric medium from the previous expressions. Assuming that the medium has dielectric function equal to $\varepsilon_{\bot}=\varepsilon_{\parallel}=\varepsilon>0$, the condition \myref{d.8} results in the canonical relation for Cherenkov radiation: $v>c/\sqrt{\varepsilon}$. Moreover, the two wavevector solutions $\vect{k}^{(1)}, \vect{k}^{(2)}$ given by Eqs. \myref{d.3.a} and \myref{d.3.b} result in
\begin{subequations}
\label{d.9}
\begin{align}
\label{d.9.a}
\vect{k}^{(1)}&=\frac{\omega}{v} \mathbb{M} \, (\uvect{y} - \sqrt{\Delta}\, \uvect{z}),\\
\label{d.9.b}
\vect{k}^{(2)}&=\frac{\omega}{v} \mathbb{M} \, (\uvect{y} + \sqrt{\Delta}\, \uvect{z}),
\end{align}
\end{subequations}
with
\begin{equation}
\label{d.10}
\mathbb{M}=\left[
\begin{array}{cc}
\sin\alpha & -\cos\alpha \\
\cos\alpha & \sin\alpha
\end{array}
\right]. 
\end{equation}
It is worthwhile noting that $\mathbb{M}$ is an orthogonal matrix.  This implies that the angles $\theta^{(1)}_{\vect{k}}$ and $\theta^{(2)}_{\vect{k}}$ are always equal. Thus, the propagation of the wake patterns excited in an isotropic dielectric media is always cylindrically symmetric with respect to the electron beam trajectory.

%
\section{Momentum-resolved loss and EEL probabilities for electron trajectories oblique to the optical axis of h-BN\label{appendE}}
%
%
\begin{figure*}
\begin{center}
\includegraphics[width=\textwidth, scale=0.83]{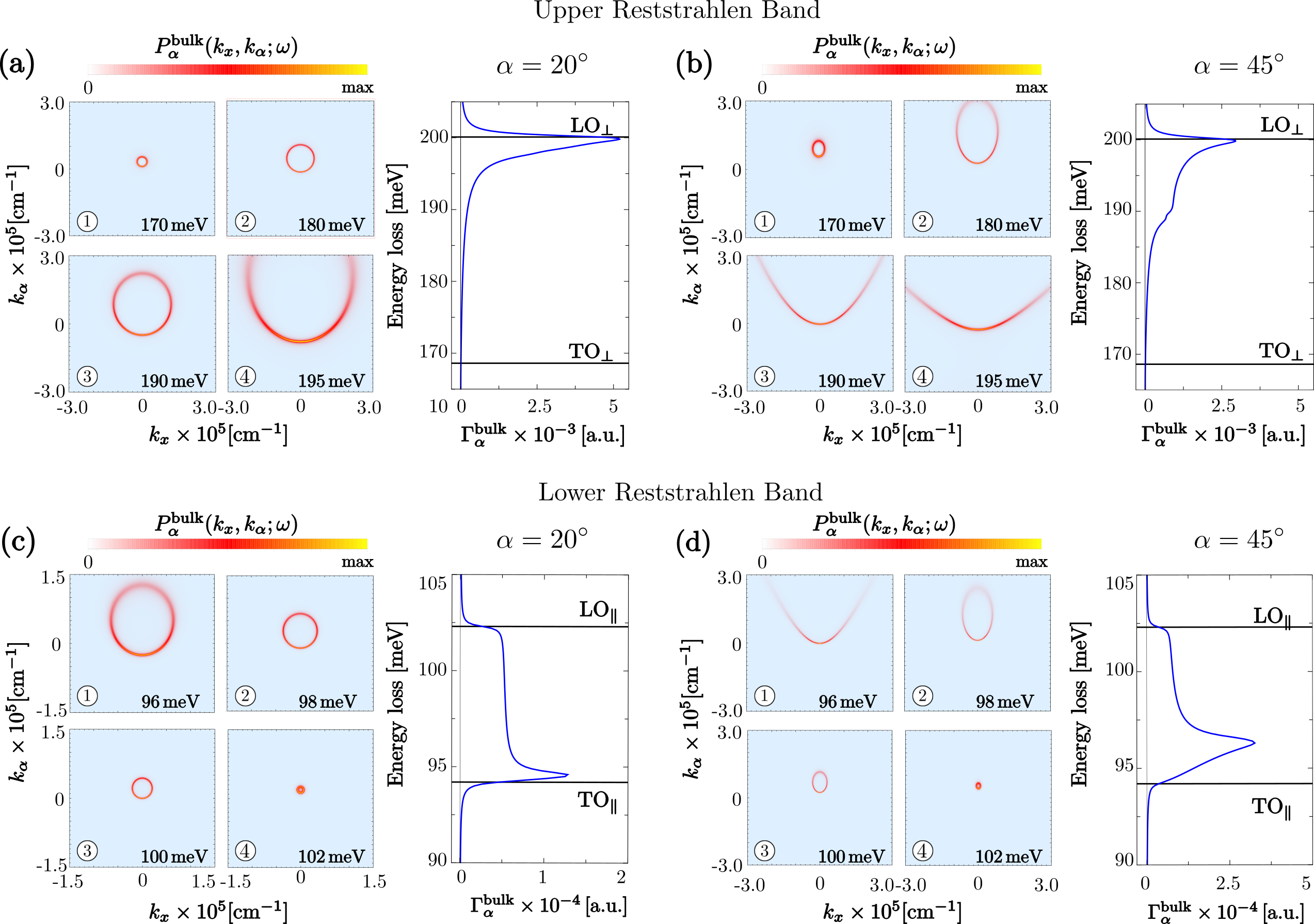}
\caption{The color plots in (a) and (b) show the momentum-resolved loss probabilities $P^{\text{bulk}}_{\alpha}(k_x,k_{\alpha};\omega)$ for representative energies within the upper Reststrahlen band (170,180,190 and 195\,meV) when the angle $\alpha$ of the electron beam trajectory is equal to (a) $20^{\circ}$ and (b) $45^{\circ}$ with $v=0.1c$.  The right panels in (a) and (b) show $\Gamma^{\text{bulk}}_{\alpha}(\omega)$ obtained by integrating  $P^{\text{bulk}}_{\alpha}(\vect{q};\omega)$ over the reciprocal coordinates $(q,\phi)$ (given by Eq. \myref{d.4}) up to the cutoff value $q^c=0.05\,\mathring{\text{A}}^{-1}$. (c) and (d) are the analogous of (a) and (b) but for representative energies with in the lower Reststrahlen band (96, 98, 199 and 102\,meV). The color plots are normalized with respect to the maximum value in each case: (a.1) 500\,a.u., (a.2) 1500\,a.u., (a.3) $>1500\,\text{a.u.}$, (a.4) 1250\,a.u.; (b.1) $>300\,\text{a.u.}$, (b.2) 2000\,a.u., (b.3) $>2500\,\text{a.u.}$, (b.4) $>2500\,\text{a.u.}$; (c.1) 1250\,a.u., (c.2) 4000\,a.u., (c.3) $>6000\,\text{a.u.}$, (c.4) $>10000\,\text{a.u.}$; (d.1) $>2000\,\text{a.u.}$, (d.2) $>5000\,\text{a.u.}$, (d.3) $>8000\,\text{a.u.}$, (d.4) $>8000\,\text{a.u.}$}
\label{fig13}
\end{center}
\end{figure*}
%
%
As we show in appendix \ref{appendD}, the cylindrical symmetry in the propagation of the phonon polariton wave is broken when the electron beam trajectory is not parallel to the h-BN optical axis. This break in symmetry means that the momentum-resolved loss probability $P^{\text{bulk}}(\vect{q};\omega)$ is no longer constant along the azimuthal direction but it depends on the angle $\phi$ \cite{Fossard2017}. Notice also that the momentum $\hbar \vect{q}=(\hbar k_x,\hbar k_y)$ is no longer perpendicular to the beam trajectory $\vect{r}_e(t)= v t (0,\sin\alpha,\cos\alpha)$. In fact, the two orthogonal directions to $\vect{r}_e(t)$ are: (i) the \textit{x}-direction and (ii) the direction set by the unit vector $\uvect{n}_{\alpha}=(0,\cos\alpha,-\sin\alpha)$. Thus, the two transverse components (to the beam trajectory) of the polariton wavevector are $k_x$ and
\begin{equation}
\label{e.1}
k_{\alpha}=\vect{k}\cdot \uvect{n}_{\alpha}=k_y \cos\alpha-k_z \sin\alpha.
\end{equation}
Furthermore, the components of the polariton wavevector $\vect{k}(\omega)$ excited by the fast electron beam need to satisfy Eq. \myref{d.1.b}. By solving Eqs. \myref{d.1.b} and \myref{e.1} one finds that $k_y$ and $k_z$ can be written in terms of $k_{\alpha}$ as
\begin{subequations}
\label{e.2}
\begin{align}
\label{e.2.a}
k_y&=k_{\alpha} \cos\alpha + \frac{\omega}{v} \sin\alpha,\\
\label{e.2.b}
k_z&=-k_{\alpha} \sin\alpha + \frac{\omega}{v} \cos\alpha.
\end{align}
\end{subequations}

Following Eq.~\myref{2.i}, we can define the probability for the fast electron to transfer a transverse momentum $(\hbar k_x, \hbar k_{\alpha})$ upon loosing energy $\hbar\omega$ as 
\begin{equation}
\label{e.3}
P_{\alpha}^{\text{bulk}}(k_x,k_{\alpha};\omega) = -\frac{2 e^2}{(2\pi)^3 \hbar c^2 \varepsilon_0 \cos\alpha} \text{Im} \left[\uvect{v} \cdot \tensor{\bf G}^* \cdot \uvect{v} \right],\\
\end{equation}
where  ${\tensor{\bf G}^*}={\tensor{\bf G}}(k_x,k^*_y,k^*_z)$ and $k^*_y$, $k^*_z$ are given by Eqs. \myref{e.2.a} and \myref{e.2.b}, respectively. On the other hand, the electron energy loss probability, $\Gamma^{\text{bulk}}_{\alpha}(\omega)$, can be obtained by integrating $P^{\text{bulk}}_{\alpha}(k_x,k_{\alpha};\omega)$ over the momentum coordinates (Eq. \myref{2.h}):
\begin{align}
\label{e.4}
\nonumber
\Gamma^{\text{bulk}}_{\alpha}(\omega) &=\int  \text{d}k_x \int \text{d}k_y \,  P_{\alpha}^{\text{bulk}}(k_x,k_y;\omega)\\
&= \int  \text{d}k_x \int \text{d}k_{\alpha} \, \cos\alpha P_{\alpha}^{\text{bulk}}(k_x,k_{\alpha};\omega)\\
\nonumber
&=\int_{0}^{q^c}  q\, \text{d}q \int_0^{2\pi} \text{d}\phi \,  P_{\alpha}^{\text{bulk}}(q,\phi;\omega),
\end{align}
where the last equality follows by expressing $\vect{q}$ in cylindrical coordinates. Notice that the integration over the magnitude of $\vect{q}$ is performed up to the cutoff value $q^c$. 

In Fig. \ref{fig13} we show the momentum-resolved loss probability $P_{\alpha}^{\text{bulk}}(k_x,k_{\alpha};\omega)$ and the EEL probability $\Gamma^{\text{bulk}}_{\alpha}(\omega)$ for representative energies inside the Reststrahlen bands when $v=0.1c$ and two different trajectory angles $\alpha$: $20^{\circ}$ and $45^{\circ}$. One can observe in the figure that the EEL features are similar but the momentum-resolved loss probability shows asymmetries for different energies in the Reststrahlen bands.
%
%
\section{Induced electromagnetic field for an electron trajectory above the surface of an uniaxial anisotropic semi-infinite medium\label{appendF}}
To obtain the induced electromagnetic field when the electron is traveling above the surface of an anisotropic media, we solve the following wave equation (derived from Maxwell's equations) satisfied by the total electric field \cite{Jackson3ed} 
\begin{align}
\label{f.1}
&\nabla^2 \vect{E}^{\text{tot}} (\vect{r};t) - \mu_0 \epsilon_0 \frac{\partial^2}{\partial t^2} [{\tensor{\epsilon}} \vect{E}^{\text{tot}}  (\vect{r};t)] =\\
\nonumber
& \mu_0 \frac{\partial}{\partial t} \vect{J}(\vect{r};t) + \nabla[\nabla \cdot \vect{E}^{\text{tot}} (\vect{r};t)],
\end{align}
where $\varepsilon_0$ and $\mu_0$ stand for the vacuum permittivity and permeability, respectively, and  $\vect{J}(\vect{r};t)=\rho(\vect{r};t) \vect{v}=-e\delta(x-x_0,0,z-vt)(0,0,v)$ is the current density corresponding to the electron traveling with velocity $\vect{v}=v \uvect{z}$ and impact parameter $x_0$. We show in Fig. 7 of the main text a schematics of the considered geometry. 

By Fourier transforming Eq. \myref{f.1} with respect to the variables $y$, $z$ and $t$ and solving for the electric field separately outside (label I) and inside (label II) the anisotropic medium, we obtain the following solutions for the components of the total electric field
\begin{subequations}
\label{f.2}
\begin{align}
\label{f.2.a}
&E^{(\text{I})}_x (x,k_y,k_z;\omega)= B_{\text{I}} \,e^{-\kappa_{\text{I}} x} \\
\nonumber
&\hspace{4mm} -  \frac{\pi e}{\varepsilon_0} \text{sign}(x-x_0) \delta(\omega-k_z v) e^{-\kappa_{\text{I}}  \lvert x-x_0 \rvert}, \\
\label{f.2.b}
&E^{(\text{I})}_y (x,k_y,k_z;\omega) = D_{\text{I}} \,e^{-\kappa_{\text{I}}  x} \\
\nonumber
&\hspace{4mm} - i \frac{\pi e}{\varepsilon_0} \frac{k_y-\frac{\omega}{c^2}v_y}{\kappa_{\text{I}}} \delta( \omega-k_z v) e^{-\kappa_{\text{I}}  \lvert x-x_0 \rvert},\\
\label{f.2.c}
&E^{(\text{I})}_z (x,k_y,k_z;\omega)  = G_{\text{I}} \,e^{-\kappa_{\text{I}} x} \\
\nonumber
&\hspace{4mm} - i \frac{\pi e}{\varepsilon_0} \frac{k_z - \frac{\omega}{c^2} v_z}{\kappa_{\text{I}} } \delta(\omega-k_z v) e^{-\kappa_{\text{I}}  \lvert x-x_0 \rvert}, \\
\label{f.2.d}
&E^{(\text{II})}_x (x,k_y,k_z;\omega) = A_{\text{II}} e^{\kappa^{o}_{\text{II}}  x} -i F_{\text{II}} \frac{k_z \kappa^{e}_{\text{II}}}{(\kappa^{o}_{\text{II}})^2 - k^2_y} e^{\kappa^{e}_{\text{II}} x}, \\
\label{f.2.e}
&E^{(\text{II})}_y (x,k_y,k_z;\omega)   = C_{\text{II}} e^{\kappa^{o}_{\text{II}} x} + F_{\text{II}} \frac{k_y k_z}{(\kappa^{o}_{\text{II}})^2 - k^2_y} e^{\kappa^{e}_{\text{II}} x}, \\
\label{f.2.f}
&E^{(\text{II})}_z (x,k_y,k_z;\omega)  = F_{\text{II}} e^{\kappa^{e}_{\text{II}} x},
\end{align}
\end{subequations}
where
\begin{align}
\label{f.3}
\nonumber
&\kappa_{\text{I}}^2= k^2_y+k^2_z - \frac{\omega^2}{c^2}, \,\, (\kappa^{e}_{\text{II}})^2= k_y^2 + \frac{\varepsilon_{\parallel}}{\varepsilon_{\bot}}\left(k^2_z - \frac{\omega^2}{c^2} \varepsilon_{\bot}\right)\\
&\text{and} \hspace{4mm} (\kappa^{o}_{\text{II}})^2= k^2_y+k^2_z -\varepsilon_{\bot} \frac{\omega^2}{c^2} .
\end{align}
The coefficients $A_{\text{II}},B_{\text{I}},C_{\text{II}},D_{\text{I}},F_{\text{II}}$ and $G_{\text{I}}$ can be found from the application of the standard  boundary conditions for the field at the interface ($x=0$) between both media, that is,
\begin{align}
\label{f.4}
&E^{(\text{II})}_y\rvert_{x=0}=E^{(\text{I})}_y\rvert_{x=0}, \hspace{5mm} E^{(\text{II})}_z\rvert_{x=0}=E^{(\text{I})}_z\rvert_{x=0},\\
\nonumber
&\varepsilon_{\bot} E^{(\text{II})}_x\rvert_{x=0}=E^{(\text{I})}_x\rvert_{x=0}.
\end{align}

%
\section{Momentum-resolved loss and EEL probabilities for electron trajectories above the surface of h-BN parallel to the optical axis\label{appendG}}
By solving the linear system of equations set by the boundary conditions (Eq. \myref{f.4}), one finds that each coefficient in Eqs. \myref{f.2.a}-\myref{f.2.f} can be expressed as
\begin{align*}
A_{\text{II}}=\tilde{\rho}\, a_{\text{II}}, \hspace{4mm} B_{\text{I}}=\tilde{\rho} \, b_{\text{I}}, \hspace{4mm} C_{\text{II}}=\tilde{\rho}\, c_{\text{II}} \\
D_{\text{I}}=\tilde{\rho}\, d_{\text{I}}, \hspace{4mm} F_{\text{II}}=\tilde{\rho} f_{\text{II}}, \hspace{4mm} G_{\text{I}}=\tilde{\rho} g_{\text{I}},
\end{align*} 
with $\tilde{\rho}=-2\pi e \delta(\omega-k_z v) e^{-\kappa_{\text{I}} x_0}/\varepsilon_0$. Thus, we obtain that the induced electric fields in vacuum (labeled as I) and h-BN (labeled as II) are given by (Eqs. \myref{f.2.a}-\myref{f.2.f})
\begin{subequations}
\label{g.1}
\begin{align}
\label{g.1.a}
&\vect{E}^{\text{ind}}_{\text{I}} (x,k_y,k_z;\omega)=(b_{\text{I}},d_{\text{I}},g_{\text{I}})\,\tilde{\rho}\,e^{-\kappa_{\text{I}} x},\\
\label{g.1.b}
&\vect{E}^{\text{ind}}_{\text{II}} (x,k_y,k_z;\omega)=(a_{\text{II}},c_{\text{II}},0)\,\tilde{\rho}\,e^{\kappa^{o}_{\text{II}} x}\\
\nonumber
&+ \left(-i \frac{k_z \kappa^{e}_{\text{II}}}{(\kappa^{o}_{\text{II}})^2-k^2_y}, \frac{k_z k_y}{(\kappa^{o}_{\text{II}})^2-k^2_y}, 1 \right)\,\tilde{\rho}\,\text{f}_{\text{II}}\,e^{\kappa^{e}_{\text{II}} x}.
\end{align}
\end{subequations}  

Substituting Eq. \myref{g.1.a} into Eq. \myref{2.f}, one obtains that the EEL probability $\Gamma^{\text{surf}}(\omega)$ can be written as
\begin{align}
\label{g.1}
\Gamma^{\text{surf}}(\omega)&= \frac{e}{\pi \hbar \omega} \text{Re} \left[ \vect{E}^{\text{ind}}_{\text{I}}({\vect{r}}_{e};\omega) \cdot \uvect{z} \, e^{-i\omega t_{e}} \right]\\
\nonumber
&=\int_0^{k^c_y} \text{d}k_y \, P^{\text{surf}}(k_y;\omega),
\end{align}
with $\hbar k^c_y$ the maximum momentum of the electrons that can pass through the collection aperture of the detector in the \textit{y}-direction, and
\begin{equation}
\label{g.2}
P^{\text{surf}}(k_y;\omega)=-\frac{e^2}{\pi^2\varepsilon_0\hbar \omega v} \text{Re}\left[g_I  e^{-2\kappa_{\text{I}} x_0} \right]\Big\lvert_{k_z=\omega/v},
\end{equation}
where $\hbar k_z=\hbar \omega/v$ is the momentum transferred by the electron to the polaritons along the beam trajectory.
%
%
%
%
\begin{figure*}[!ht]
\begin{center}
\includegraphics[scale=0.83]{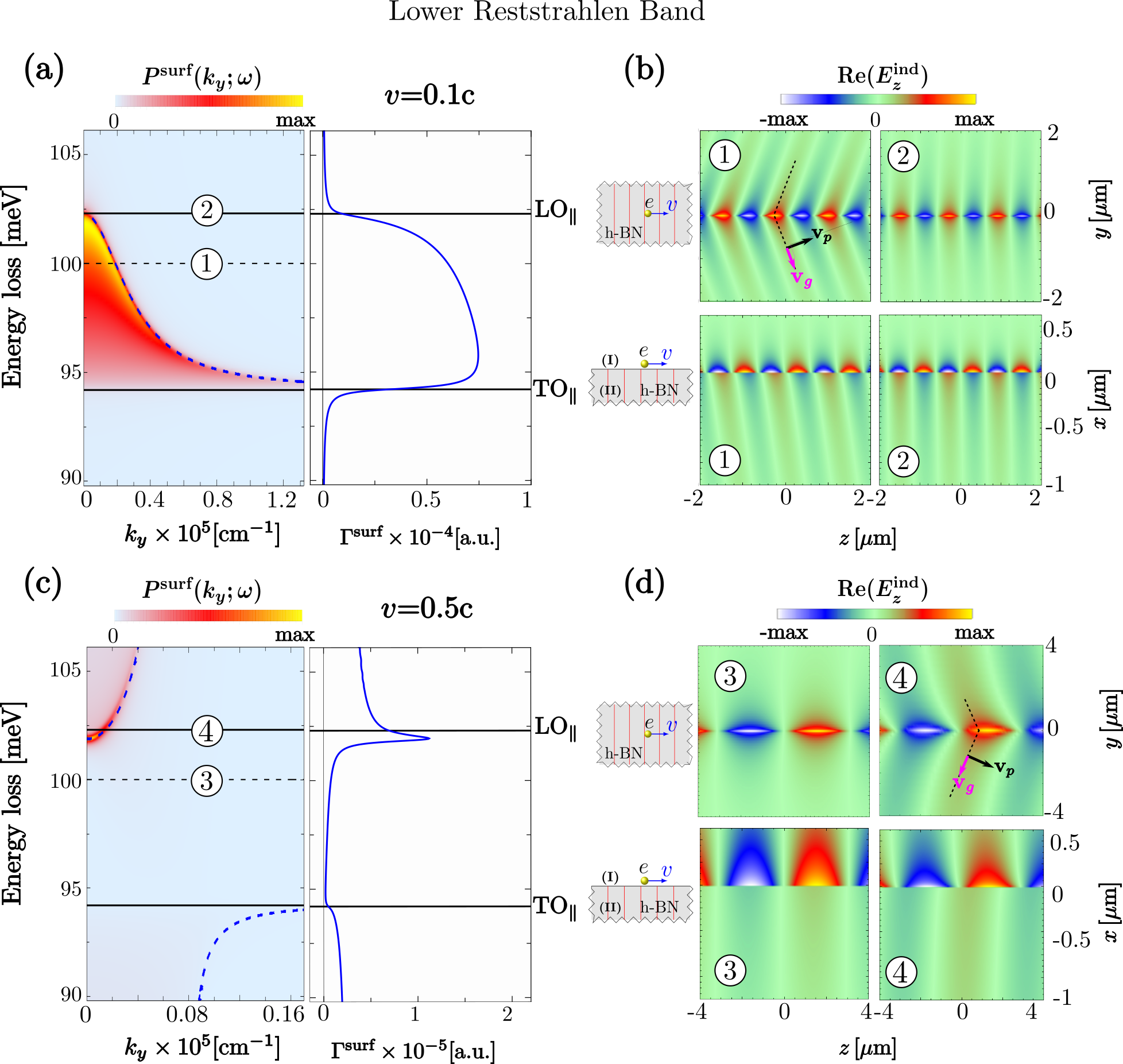}
\caption{The left panel in (a) displays the momentum-resolved loss probability $P^{\text{surf}}(k_y;\omega)$ normalized to the maximum value ($>0.6\,\text{a.u.}$) in the vicinity of the lower Reststrahlen band for $x_0=10\,\text{nm}$ and $v=0.1c$. The right panel in (a) shows the EEL probability $\Gamma^{\text{surf}}(\omega)$ obtained by integrating $P^{\text{surf}}(k_y;\omega)$ over $k_y$ up to $k^c_{y}=0.09\,\mathring{\text{A}}^{-1}$. (c) same as in (a) but considering $v=0.5c$. For this case the maximum value of the momentum-resolved loss probability is 2\,a.u. The color maps in (b) and (d) show the real part of the \textit{z}-component of the induced electric field for the energies: 100 (marked 1, 3) and $\text{LO}_{\parallel}$ (marked 2, 4). The top panels in (b) and (d) correspond to the in-plane views ($yz$ plane) of the induced field, while the bottom panels correspond to the out-of-plane views ($xz$ plane).  The field plots are normalized with respect to the maximum value in each case. For the top panels: (b) $7.5\times10^{-6}\,\text{a.u.}$, (d.3) $4\times10^{-7}\,\text{a.u.}\,$  and (d.4) $3\times10^{-7}\,\text{a.u.}$ For the bottom panels: (b) $5\times10^{-6}\,\text{a.u.}$ and (d.3) $3\times10^{-7}\,\text{a.u.}$, (d.4) $4\times10^{-7}\,\text{a.u.}$}
\label{fig14}
\end{center}
\end{figure*}
%
%
\section{Electron energy loss probability for  energies around the lower Reststrahlen band for electron trajectories above the surface of h-BN\label{appendH}}
In the  left panel of Fig. \ref{fig14}a we show the momentum-resolved loss probability, $P^{\text{surf}}(k_y;\omega)$, for an electron traveling above an h-BN surface for energies around the lower Reststrahlen band. The probing electron travels above the surface at an impact parameter of 10\,nm and $v=0.1c$. The blue dashed line corresponds to the bulk phonon polariton dispersion (Eq. \myref{2.c}). We can recognize some similarities between $P^{\text{surf}}(k_y;\omega)$ and $P^{\text{bulk}}(k_{\bot};\omega)$ (compare the left panels of  Figs. \ref{fig4}b and \ref{fig14}a). For instance, the maximum values of $P^{\text{surf}}(k_y;\omega)$  are close to the bulk dispersion (blue dashed line). Interestingly, this bulk dispersion corresponds to the envelope curve of $P^{\text{surf}}(k_y;\omega)$ implying that electron energy losses in the lower band are mainly due to bulk hyperbolic phonon polariton excitations. To obtain spectroscopic information on the excitations in the lower band, we calculate the EEL probability $\Gamma^{\text{surf}}(\omega)$ by integrating $P^{\text{surf}}(k_y;\omega)$ over $k_y$ up to a cutoff $k_y^c$ (right panel in Fig. \ref{fig14}a). Similarly to $\Gamma^{\text{bulk}}(\omega)$ (Fig. \ref{fig4}b, right panel), $\Gamma^{\text{surf}}(\omega)$ (right panel in Fig. \ref{fig14}a) exhibits a uniform loss probability between $\text{TO}_{\parallel}$ and $\text{LO}_{\parallel}$ which depends on the selected cutoff momenta $\hbar k_y^c$. 

In Fig. \ref{fig14}b we show the real part of the \textit{z}-component of the induced electric field for the same electron velocity and impact parameter as in Fig. \ref{fig14}a, for two different energies marked as 1 and 2 in panel a.  We can recognize the excitation of the wake fields in the h-BN surface for those energy losses (compare the top panels labeled as 1 and 2 in Fig \ref{fig14}b). The bulk nature of the excited modes is revealed in the bottom panels of Fig. \ref{fig14}b, where we show \textit{z}-component of the real part of the induced electric field $\text{Re}(E^{\text{ind}}_z(\vect{r};\omega))$ in the $xz$-plane. In this lateral view of the field distribution one notice the excitation and propagation of the field into the bulk from the h-BN surface.  

Panels \ref{fig14}c and \ref{fig14}d show $P^{\text{surf}}(k_y;\omega)$, $\Gamma^{\text{surf}}(\omega)$ and the induced field distribution when the electron velocity is $0.5c$. It is worth noting that the blue dashed line superimposed on $P^{\text{surf}}(k_y;\omega)$ (Fig. \ref{fig14}c, left panel) corresponds to another branch of Dyakonov's dispersion relation given by Eqs. \myref{3.c.a}-\myref{3.c.c} and \myref{3.d}.
%
%

%
\end{document}